\documentclass[aps,pra,reprint,superscriptaddress]{revtex4-2}
\usepackage{lipsum}
\usepackage{graphicx}
\usepackage{amsmath}
\usepackage[colorlinks=true, linkcolor=blue, citecolor=blue, urlcolor=blue]{hyperref}

\begin{document}
\title{Nonadiabatic theory for subcycle ionic dynamics in multielectron tunneling ionization}
\author{Chi-Hong Yuen}
\email[]{cyuen2@kennesaw.edu}
\affiliation{Department of Physics, Kennesaw State University, Marietta, GA 30060, USA}

\date{\today}

\begin{abstract}
Multielectron tunneling ionization creates ionic coherence crucial for lasing and driving electron motion in molecules.
While tunneling is well understood as a single active electron process, less emphasis has been placed on theoretical descriptions of bound electrons during tunneling.
This work systematically investigates multielectron tunneling ionization based on the strong field approximation, establishing a theoretical foundation and demonstrating the equivalence of wave function and density matrix approaches for subcycle ionic dynamics.
An accurate subcycle nonadiabatic ionization rate is also derived and incorporated into the theory to improve its quantitative accuracy.
Applying the theory to N$_{2}$ and CO$_{2}$, this work showcases how an intense laser field can induce ionic coherence in molecules as observed in previous experiments.
These findings encourage future investigations into multielectron tunneling ionization and its applications in lasing and in controlling chemical reactions.
\end{abstract}

\maketitle

\section{Introduction~\label{sec1}}
Multielectron tunneling ionization of molecules has attracted significant interest owing to its relevance to the control of ultrafast chemical processes and lasing.
Tunneling ionization has been intuitively understood as a single active electron process based on the strong field approximation (SFA)~\cite{Keldysh1965, Popruzhenko2014}.
When a molecule is exposed to an intense laser field, the valence electron in an occupied molecular orbital encounters a potential barrier and can tunnel out, thereby becoming ionized.
Substantial success has been achieved by applying the SFA in the single active electron picture to strong field phenomena~\cite{Amini2019}.
However, the multielectron nature of molecules means that an electron can be tunnel ionized from multiple valence orbitals to create different ionic states~\cite{McFarland2008, Smirnova2009, Akagi2009, Haessler2010, Farrell2011, De2011, Boguslavskiy2012, Wu2012, Kotur2012} which can be strongly coupled by the laser field~\cite{Kraus2015, Kobayashi2020, Kobayashi2020a, Lei2022, Kleine2022, He2022, Shu2022, Chen2024, Gao2025, Gao2025co2}, forming a superposition of ionic states.
The electronic coherence in the ion leads to electron motion within the molecule, or charge migration, as observed in high harmonic spectroscopy~\cite{Smirnova2009, Haessler2010, Mairesse2010, Kraus2015, He2022, He2023}.
When nuclear motion sets in on different ionic potentials, the electronic coherence may undergo dephasing or decoherence~\cite{Kobayashi2020, Kobayashi2020a}, potentially leading to charge transfer.
Therefore, controlling the electronic coherence of the molecular ion created by an intense laser field is closely linked to attochemistry~\cite{Lepine2014, Merritt2021, Calegari2023}.
On the other hand, vibronic coherence in the ion enables coherent light emission, leading to air lasing~\cite{Yao2011, Liu2013, Liu2015, Xu2015, Yao2016, Chen2024, Gao2025, Xu2025} and the generation of broadband ultraviolet (UV) light~\cite{Lei2022, Gao2025co2}.
Given the wide-ranging applications of ionic coherence generated by strong laser fields, there is motivation to understand how it develops from a multielectron perspective.

Subcycle ionic dynamics in multielectron tunneling ionization has been studied using wave function approaches.
In high harmonic spectroscopy, the ionic wave function was propagated from the birth time to the return time of the ionized electron by solving the time-dependent Schr\"{o}dinger equation (TDSE), while the tunneling of the active electron was modeled using the Yudin-Ivanov formula~\cite{Yudin2001, Smirnova2009, Mairesse2010}, the weak field asymptotic theory~\cite{Tolstikhin2011, Kraus2015}, or the molecular Ammosov-Delone-Krainov (ADK) theory~\cite{Tong2002, He2022, He2023}.
The measured spectra were used to reconstruct the attosecond electron wave packet of the ion based on the theory.
On the other hand, in modeling air lasing, only the direct electrons are considered.
Therefore, the ionic wave function should be propagated from the birth time to the end of the laser pulse.
It was found that population inversion between the $B^{2}\Sigma_{u}^{+}$ and $X^{2}\Sigma_{g}^{+}$ states of N$_{2}^{+}$ could occur~\cite{Xu2015, Zhang2017, Xu2022} and thereby explain the mechanism of air lasing.
While wave function approaches have successfully explained experimental measurements, one might question their validity when the ionized electron is driven solely by the strong laser field and drifts away from the ion, suggesting that the ion should be treated as an open quantum system.

The native approach for the evolution of an open quantum system is the density matrix formalism.
A density matrix is first constructed from the total wave function, and then the active electron part is traced out.
The ionic dynamics can then be determined by evolving its reduced density matrix.
Santra and coworkers first introduced the use of the density matrix in first-principles calculations to describe strong-field-generated ionic coherence in noble gas atoms with spin-orbit couplings~\cite{Rohringer2009, Santra2011}.
The theory was used to simulate the attosecond transient absorption spectrum and to reconstruct the motion of valence electrons in the krypton ion in the seminal experiment by Goulielmakis \textit{et al.}~\cite{Goulielmakis2010}.
Other density matrix approaches for strong field phenomena have been based on the ADK theory~\cite{Ammosov1986, Tong2002}.
Pfeiffer \textit{et al.} developed a density matrix approach for sequential double ionization of noble gas atoms with spin-orbit couplings~\cite{Pfeiffer2013}.
It was used to interpret and simulate the attosecond transient absorption spectrum of xenon dication created by a strong laser field~\cite{Kobayashi2018}.
The ionization-coupling theory was developed by Zhang \textit{et al.}~\cite{Zhang2020} to investigate the mechanism of lasing in N$_{2}^{+}$.
Their model also explains the generation of supercontinuum UV light~\cite{Lei2022} and wavelength-dependent enhancement in air lasing~\cite{Chen2024}.
Additionally, Yuen and coworkers developed a density matrix approach for sequential double or triple ionization of N$_{2}$~\cite{Yuen2022, Jia2024, Jia2025} and O$_{2}$~\cite{Yuen2023}, with good agreement with the experiments on the kinetic energy release spectra of fragmented ions~\cite{Voss2004, Wu2010, Wu2010a}.
Their approach also explains the measurements in strong field electronic coherence spectroscopy~\cite{Yuen2024a, Weckwerth2025} and the carrier-envelope phase dependence of ionic coherence in N$_{2}^{+}$~\cite{Gao2025}.

Since both wave function and density matrix approaches can well explain measurements in strong field experiments, the question arises: when rescattering can be neglected, is it necessary or advantageous to trace out the active electron from the ion-electron system to model ionic dynamics?
Additionally, the equations of motion in the density matrix approach were derived intuitively rather than formally~\cite{Yuen2022, Yuen2023, Yuen2023b, Yuen2024b, Jia2024, Jia2025}.
What is the theoretical basis for such a density matrix approach?
Can the quantitative accuracy of these equations be improved?
This article aims to address these questions systematically.

This article is structured as follows.
In the next section, we briefly review the SFA in the single active electron picture and derive the subcycle nonadiabatic ionization rate.
In Sec.~\ref{sec3}, we extend the SFA to the multielectron picture and derive the total wave function during a half-cycle.
We proceed to examine the longitudinal canonical momentum and accumulated phases of the active electrons from different occupied molecular orbitals in Sec.~\ref{sec4}.
The density matrix approach for multielectron tunneling ionization is derived in Sec.~\ref{sec5} and its applications to molecules are presented in Sec.~\ref{sec6}.
Finally, in Sec.~\ref{sec7}, we present the conclusion from this work.
Atomic units are used throughout this article unless specified.

\section{Single Active Electron Approach~\label{sec2}}
To keep this article self-contained, we briefly review the single active electron SFA here.
Detailed discussion or review can be found in Ivanov \textit{et al.}~\cite{Ivanov2005} or Popruzhenko~\cite{Popruzhenko2014}.
We partition the full Hamiltonian of the active electron as $\hat{H} = \hat{H}_{0} + \hat{V}_{L}$, where $\hat{H}_{0}$ is the field-free Hamiltonian and $\hat{V}_{L} (t)= - \mathbf{d} \cdot \mathbf{F}(t)$ is the interaction between the laser $\mathbf{F}(t) = F_{0} \cos \omega t \, \mathbf{\hat{z}}$ and the active electron in the length gauge.
Denoting the initial wave function as $|\Psi_{0} \rangle$, the formal solution to the TDSE is
\begin{align}
|\Psi(t) \rangle &= -i \int_{t_i}^{t} dt' e^{-i \int_{t'}^{t} \hat{H}(t'')dt''} \hat{V}_{L}(t') e^{-i \int_{t_i}^{t'} \hat{H_{0}}(t'')dt''} \nonumber \\
&\times |\Psi_{0} \rangle + e^{-i \int_{t_i}^{t} \hat{H_{0}}(t'')dt''} |\Psi_{0} \rangle.
\label{eq:TDSE}
\end{align}
In the single active electron picture, $|\Psi_{0} \rangle$ is taken as the wave function of the ionizing orbital with an ionization potential $E$.
The amplitude for an ionized electron with canonical momentum $\mathbf{p} = (p_{\perp}, p_{z})$ at birth time $t$ is~\cite{Ivanov2005}
\begin{align}
a_{\mathbf{p}}(t) &= -i \int_{t_i}^{t} dt' \langle \mathbf{p} + \mathbf{A}(t) | e^{-i \int_{t'}^{t} \hat{H}(t'')dt''} \hat{V}_{L}(t') \nonumber \\
&\times e^{-i \int_{t_i}^{t'} \hat{H_{0}}(t'')dt''} |\Psi_{0} \rangle \label{eq:SAE_amp} \\
&\approx -i \int_{t_i}^{t} dt'  e^{-i \left\{ \frac{1}{2}  \int_{t'}^{t}  [\mathbf{p} + \mathbf{A}(t'')]^{2} dt'' - E (t' - t_i) \right\}} \nonumber \\
&\times \langle\mathbf{p} + \mathbf{A}(t') | \hat{V}_{L}(t') |\Psi_{0} \rangle
\label{eq:SAE_amp2}
\end{align}
where $|\mathbf{p} + \mathbf{A}(t)\rangle$ is the Volkov wave function in the length gauge at time $t$.
In the SFA, the Hamiltonian $\hat{H}$ for the ionized electron is approximated as $\hat{H}_{F}=\hat{H} - \hat{V}_{0}$, where $\hat{V}_{0}$ is the field-free potential energy.
We then have
\begin{align}
a_{\mathbf{p}}(t) &\approx -i \int_{t_i}^{t} dt' C(t') e^{-i S_{\mathbf{p}}(t, t')}, \\
S_{\mathbf{p}}(t, t') &= \frac{1}{2} \int_{t'}^{t} dt'' [\mathbf{p} + \mathbf{A}(t'')]^{2} - E t', \label{eq:action} \\
\mathbf{A}(t) &= -\frac{F_{0}}{\omega} \sin \omega t \, \mathbf{\hat{z}}, \label{eq:vectorpot}
\end{align}
where $S_{\mathbf{p}}$ is the classical action, $\mathbf{A}$ is the vector potential of the laser field, and $C$ is the normalization factor for the ionization rate that will be determined later.

Since the classical action $S_{\mathbf{p}}(t, t')$ is sharply peaked and oscillates rapidly over $t'$, the amplitude $a_{\mathbf{p}}(t)$ can be approximated using the saddle point method.
The saddle point $t_{s}$ is found by solving
\begin{align}
\frac{\partial S_{\mathbf{p}}}{\partial t'} = \frac{1}{2} [\mathbf{p} + \mathbf{A}(t_{s})]^{2} + E = 0.
\label{eq:saddle}
\end{align}

The complex saddle point solution $t_{s}$ is the moment when the electron enters the classical forbidden region under the barrier~\cite{Ivanov2005}.
The instantaneous tunneling picture~\cite{Eckle2008, Torlina2015, Han2019, Sainadh2019} implies that $t_{s} = t + i \tau$, with $\tau$ being the tunneling time of the electron.
Consequently, this condition enforces the longitudinal exit velocity $v_{z}$ to be a function of time $t$ and transverse canonical momentum $p_{\perp}$.
For time $t$ in $(-\pi/2\omega, \pi/2\omega)$, the saddle point equation has been solved analytically by Li \textit{et al.}~\cite{Li2016},
\begin{align}
\sinh \omega \tau &= \gamma (p_{\perp}, t), \label{eq:tau} \\
\gamma (p_{\perp}, t)&=  \frac{\omega \sqrt{\kappa^{2} + p_{\perp}^{2}} }{ |F(t)|}, \label{eq:gamma} \\
p_{z}(p_{\perp}, t) &= \frac{F_{0} \sin \omega t}{\omega} \sqrt{1 + \gamma^{2}(p_{\perp}, t)}, \label{eq:pz} \\
v_{z}(p_{\perp}, t) &= p_{z}(p_{\perp}, t) -  \frac{F_{0} \sin \omega t}{\omega}, \label{eq:vz} 
\end{align}
where $\kappa = \sqrt{2E}$.
As a result, each canonical momentum $\mathbf{p}$ can be assigned to a birth time $t_{\mathbf{p}}$ by solving Eqs.~\eqref{eq:gamma} and \eqref{eq:pz}.
To simplify the notation, hereafter, we write $\gamma_{\perp t} \equiv \gamma (p_{\perp}, t)$ and $\gamma (0, 0)$, which is the Keldysh parameter~\cite{Keldysh1965}, as $\gamma_{K}$.

The real and imaginary part of the classical action $S_{\mathbf{p}}$ at a time $t \geq t_{\mathbf{p}}$ is,
\begin{align}
\mathrm{Re}[S_{\mathbf{p}}(t, t_{\mathbf{p}})] & = \left(E + \frac{p^{2}}{2} + U_{p}\right) (t-t_{\mathbf{p}}) \nonumber \\
&+ p_{z} \frac{F_{0}}{\omega^{2}} \bigg[ \cos \omega t - \cos \omega t_{\mathbf{p}} \sqrt{1 + \gamma_{\perp t_{\mathbf{p}}}^{2}} \bigg] \nonumber \\
&- \frac{U_{p}}{2\omega} \bigg[ \sin 2\omega t - \sin 2\omega t_{\mathbf{p}} (1 + 2\gamma^{2}_{\perp t_{\mathbf{p}}}) \bigg]
\label{eq:Re-S}
\end{align}
and
\begin{align}
\mathrm{Im}[S_{\mathbf{p}}(t_{\mathbf{p}})] & = -\left(E + \frac{p^{2}}{2} + U_{p}\right) \frac{\operatorname{arcsinh} \gamma_{\perp t_{\mathbf{p}}}}{\omega} \nonumber \\
&+ p_{z} \frac{F_{0}}{\omega^{2}}  \gamma_{\perp t_{\mathbf{p}}} \sin \omega t_{\mathbf{p}} \nonumber \\
&+ \frac{U_{p}}{\omega}  \gamma_{\perp t_{\mathbf{p}}} \sqrt{1 + \gamma_{\perp t_{\mathbf{p}}}^{2}} \cos 2\omega t_{\mathbf{p}},
\label{eq:Im-S}
\end{align}
where $p^{2} = p_{\perp}^{2} + p_{z}^{2}$ and $U_{p} = F_{0}^{2} / 4\omega^{2}$ is the ponderomotive energy.
This subcycle nonadiabatic theory~\cite{Li2016} has been shown to agree qualitatively with the experiment in Ref.~\cite{Li2017}.

The amplitude for an ionized electron with canonical momentum $\mathbf{p}$ is
\begin{align}
a_{\mathbf{p}}(t) &\approx \Theta(t-t_{\mathbf{p}}) e^{-i S_{\mathbf{p}}(t, t_{\mathbf{p}})} C(t_{\mathbf{p}}),
\label{eq:a_amp}
\end{align}
where $\Theta$ is the Heaviside step function with $\Theta(0) = 1/2$.
This function is introduced to ensure the amplitude is zero before the birth time.

To obtain the normalization factor, in Appendix~\ref{appendix1}, we calculated the cycle-averaged ionization rate by applying the saddle point method on the integral of $p_{\perp}$ and $p_{z}$, and matched it with the cycle-averaged Perelomov-Popov-Tenent'ev (PPT) rate~\cite{perelomov1966} with the Coulomb correction from Popruzhenko \textit{et al.}~\cite{Popruzhenko2008}. 
The applicability of the formula is $\omega / E \ll 1$ and $F_{0}/\kappa^{3} \ll 1$~\cite{Popruzhenko2014}.
These conditions lead to a negligible contribution from ionizing an orbital with a magnetic quantum number $|m| > 0$.
The normalization factor can be approximated by its value at $t = 0$ with a relative error of about 5\% for the ionization yield.

In Appendix~\ref{appendix1}, we show that, for an orbital with a magnetic quantum number $m=0$,
\begin{align}
|C(t=0)|^{2} &= \frac{|B_{0}|^{2}}{F_{0} \kappa^{2Z/\kappa}} 
\sqrt{\frac{g(\gamma_{K}) h(\gamma_{K})}{\pi}} \nonumber \\
& \times A_{0}(\omega, \gamma_{K}) 
\left(\frac{F_{0} \sqrt{1+\gamma_{K}^{2}}}{2 \kappa^{3}}\right)^{1/2}  \nonumber \\
& \times \left( \frac{2\kappa^{3}}{F_{0}} \right)^{2Z/\kappa}
\left( 1 + 2 e^{-1} \gamma_{K} \right)^{-2Z/\kappa}.
\label{eq:normalization}
\end{align} 

The Coulomb corrected subcycle ionization rate for an orbital with a magnetic quantum number $m=0$ is found to be
\begin{align}
w(t) &= \frac{|B_{0}|^{2}}{\kappa^{2Z/\kappa-1}} 
\left( 2h(\gamma_{K})\sqrt{1 + \gamma_{K}^{2}} \right)^{1/2} \nonumber \\
& \times  A_{0}(\omega, \gamma_{K}) 
\left(\frac{F_{0} \sqrt{1+\gamma_{K}^{2}}}{2 \kappa^{3}}\right)  \nonumber \\
& \times  \left( \frac{2\kappa^{3}}{F_{0}} \right)^{2Z/\kappa}
\left( 1 + 2 e^{-1} \gamma_{K} \right)^{-2Z/\kappa}
\exp{\left[-\frac{2\kappa^{3}}{3F_{0}} g(t)\right]}.
\label{eq:NA-gamma}
\end{align}

In the above, 
$Z$ is the charge of the residual ion and $g(t)$ is defined by Eq. \eqref{eq:ggamma} with $p_{\perp} = 0$.
The defintion of $g(\gamma_{K})$, $h(\gamma_{K})$,  $B_{0}$, and $A_{0}(\omega, \gamma_{K}) $ can be found in Eqs. \eqref{eq:g0}, \eqref{eq:hfunc}, \eqref{eq:Bm}, and \eqref{eq:Am}, respectively.

In the quasistatic limit $\gamma_{K} \ll 1$, we have
$g(\gamma_{K}),  A_{m}(\omega, \gamma_{K}), 2h(\gamma_{K}) \to 1$,
and $g(t) \to 1/|\cos \omega t|$.
With a relative error of around 5\% in the ionization yield, we can replace $F_{0}$ with $F(t)$ in the prefactor.
Then, equation~\eqref{eq:NA-gamma} becomes the subcycle ADK ionization rate~\cite{Bisgaard2004, Tong2002},
\begin{align}
w(t) &= \frac{|B_{0}|^{2}}{\kappa^{2Z/\kappa-1}} 
\left( \frac{2\kappa^{3}}{|F(t)|} \right)^{2Z/\kappa - 1} \exp{\left[-\frac{2\kappa^{3}}{3|F(t)|}\right]}.
\label{eq:ADK}
\end{align}

\begin{figure}[t]
\begin{center}
\includegraphics[width=8cm]{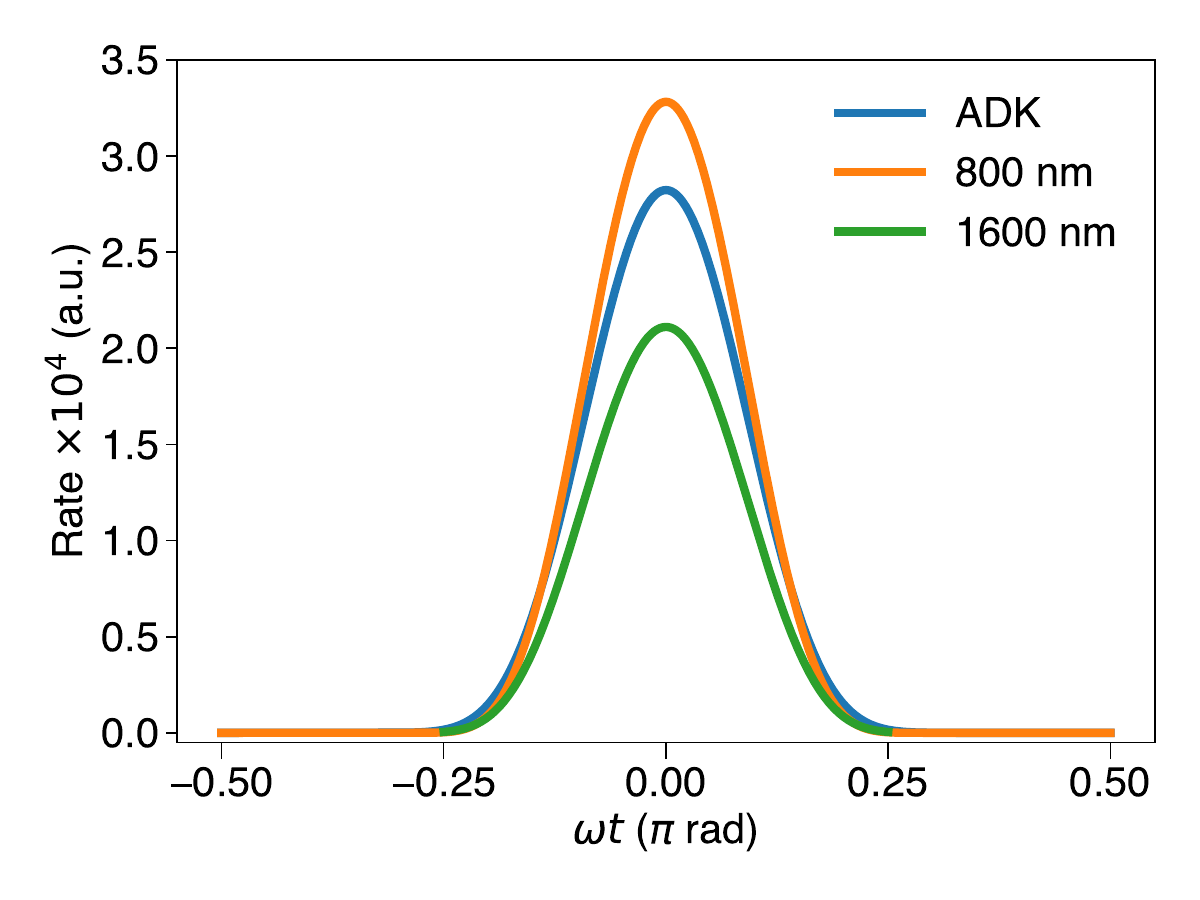}
\caption{Subcycle adiabatic (blue line) and nonadiabatic ionization rate for the H $1s$ orbital for a sinusoidal laser field with a peak laser intensity of 1.00 $\times 10^{14}$ W/cm$^{2}$ and wavelength of 800 nm (orange line) and 1600 nm (green line).}
\label{fig:rate_comp}
\end{center}
\end{figure}

Figure~\ref{fig:rate_comp} compares the subcycle ionization rate for the H $1s$ orbital calculated by Eqs.~\eqref{eq:NA-gamma} and \eqref{eq:ADK}, where $B_{0}=\sqrt{2}$ and $\kappa = 1$.
The laser field is sinusoidal with a peak intensity of 1.00 $\times 10^{14}$ W/cm$^{2}$.
When the laser wavelength is 800 nm ($\gamma_{K} = 1.07$), the nonadiabatic rate at the peak field is approximately 14\% higher than the ADK rate.
When the wavelength is 1600 nm ($\gamma_{K} = 0.53$), the nonadiabatic rate at the peak field is around 33\% lower than the ADK rate.
Interestingly, the width of the nonadiabatic rate is about the same as the ADK rate, in contrast to the Yudin-Ivanov rate~\cite{Yudin2001}.

To account for the envelope function $f(t)$ of a laser pulse, following Yudin and Ivanov~\cite{Yudin2001}, we assume it is approximately constant over half an optical cycle.
Therefore, Eqs.~\eqref{eq:NA-gamma} can be applied by replacing $F_0 \to F_0 f(t)$.
To account for the carrier-envelope phase (CEP) $\varphi$, we replace $\omega t$ in Eqs.~\eqref{eq:gamma} and \eqref{eq:ggamma} to $\omega t + \varphi$.

\begin{figure*}[t]
\begin{center}
\includegraphics[width=17cm]{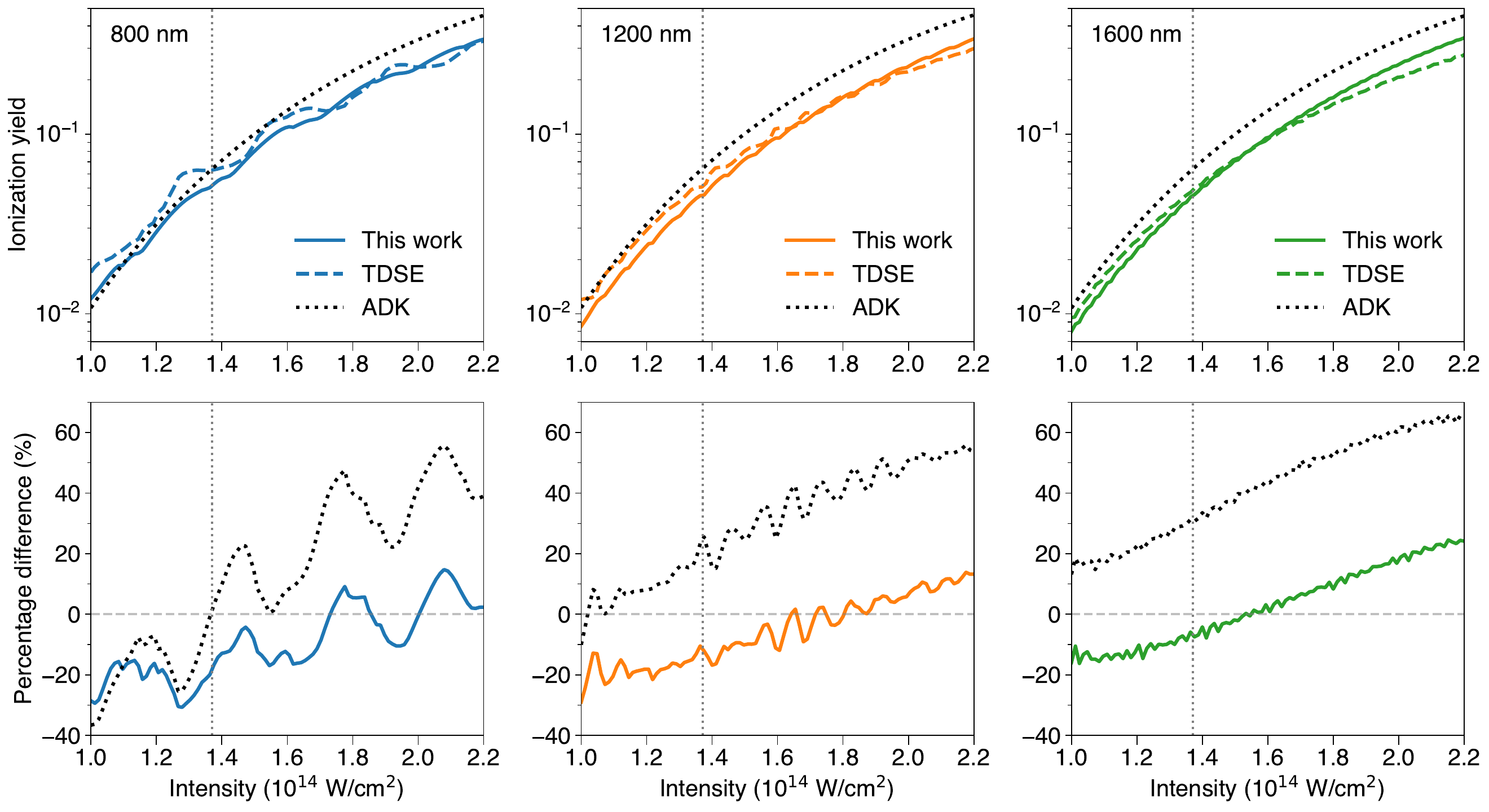}
\caption{Top row: Ionization yield for the H $1s$ orbital for a 10 fs cosine-square envelope pulse with a wavelength of 800 nm, 1200 nm, and 1600 nm at different peak laser intensities. The time-dependent Schr\"{o}dinger equation (TDSE) results are extracted from Ref.~\cite{Li2014b}. The vertical line marks the critical intensity for the H $1s$ orbital. Bottom row: The percentage difference of the yield calculated using the ADK rate and the rate in this work with the TDSE results.}
\label{fig:yield_comp}
\end{center}
\end{figure*}

To benchmark the rate from Eq.~\eqref{eq:NA-gamma} for laser pulses, we compare the ionization yield with that from solving the TDSE for the H $1s$ orbital at different wavelengths.
We follow Ref.~\cite{Li2014b} to use a cosine-square envelope function,
\begin{align}
f(t) = \cos^{2}(\pi t /\tau),
\end{align}
where the full-width-at-half-maximum (FWHM) is 10 fs and $\tau = 2.75 \times \mathrm{FWHM}$. The CEP is set to zero.
The ionization yield is calculated by 
\begin{align}
P = 1 - \exp{\left[-\int_{-\tau/2}^{\tau/2} w(t) dt\right]}.
\end{align}

Figure~\ref{fig:yield_comp} compares the intensity dependence of the ionization yield, calculated using the rate from Eqs.~\eqref{eq:NA-gamma} and \eqref{eq:ADK}, with the TDSE results extracted from Ref.~\cite{Li2014b} at wavelengths of 800, 1200, and 1600 nm.
Across all three wavelengths, the overall trend in the yield calculated using the ADK rate agrees with that from the nonadiabatic rate and the TDSE.
Below the critical intensity, the yields from the ADK rate differ by around 20 to 40\% compared to the TDSE results.
Above the critical intensity, the yields from the ADK rate can differ by up to about 60\% from those obtained from TDSE.
In contrast, the yields from the nonadiabatic rate differ by less than 30\% across the peak intensities considered.
Therefore, the nonadiabatic rate from Eq.~\eqref{eq:NA-gamma} significantly lowers the errors in the ionization yield.

On the other hand, the modulation of the ionization yield over peak intensities varies with wavelength.
For 800 nm, the TDSE yield shows large modulation over peak intensities due to the formation of the neutral excited states~\cite{Li2014b}.
There are also slight modulations in the yield from the nonadiabatic rate, but it is due to the intensity dependence of the $A_{0}(\omega, \gamma_{K})$ function \eqref{eq:Am}.
As the wavelength increases to 1600 nm, the modulation of the TDSE yield is much weaker as the excitation probability decreases significantly~\cite{Li2014b}.
At the same time, $A_{0}(\omega, \gamma_{K})$ becomes smooth at 1600 nm, and the nonadiabatic rate behaves similarly to the ADK rate.
However, at 1600 nm, the ADK rate continues to overestimate the yield at higher peak intensities.
The nonadiabatic rate improves the accuracy of the yield via the Coulomb correction factor proposed by Popruzhenko \textit{et al.}~\cite{Popruzhenko2008}.
Therefore, even for mid-infrared laser pulses, that Coulomb correction factor should not be neglected.

To summarize, we derived an analytical formula for the subcycle nonadiabatic ionization rate within the single active electron picture.
The main development was to normalize the subcycle ionization amplitude from Li \textit{et al.}~\cite{Li2016} to match the improved cycle-averaged PPT rate~\cite{perelomov1966, Popruzhenko2008} for a sinusoidal laser field, which has been known to be accurate compared to TDSE calculations~\cite{Popruzhenko2008, Zhao2016}.
Extending the formula for a laser pulse, we found that the calculated ionization yield is also accurate compared to TDSE calculations~\cite{Li2014b} over a range of peak intensities, in contrast to the yield calculated using the ADK rate.
This rate will be more versatile than the ADK rate due to its improved accuracy and broader applicability, which requires $\omega / E \ll 1$ and $F_{0}/\kappa^{3} \ll 1$, regardless of the value of the Keldysh parameter $\gamma_{K}$~\cite{Popruzhenko2014}.

\section{Multielectron approach~\label{sec3}}
In this section, we discuss the multielectron dynamics in tunneling ionization.
Tunneling has been understood as a single active electron process: Eq.~\eqref{eq:TDSE} implies that only the active electron from an occupied orbital tunnels through the potential barrier, whereas the remaining electrons are spectators.
A natural question to ask is: How to describe the dynamics of the bound electrons during and after the tunneling ionization of the active electron?

To apply the SFA to multielectron systems, we first modify Eq.~\eqref{eq:SAE_amp} as
\begin{align}
a_{i \mathbf{p}}(t) &= -i \int_{t_i}^{t} dt'  \langle i | \otimes \langle \mathbf{p} + \mathbf{A}(t) | e^{-i \int_{t'}^{t} \hat{H}(t'')dt''} \hat{V}_{L}(t') \nonumber \\
&\times c_{0}(t') e^{-i \int_{t_i}^{t'} \hat{H_{0}}(t'')dt''} |\Psi^{N+1}_{0} \rangle,
\label{eq:ME_amp}
\end{align}
where $|\Psi^{N+1}_{0} \rangle$ is the $N+1$ electrons wave function and $| i \rangle$ is the wave function for the ionic state $i$.
The amplitude $c_{0}(t')$ accounts for the depletion of the neutral state, which follows from the SFA with the dressed initial state~\cite{Smirnova2007}.
However, we neglect the other effects of the laser field on the neutral ground state for simplicity.
$\hat{H_{0}}$ and $\hat{H}(t)$ are now $N+1$ electrons Hamiltonian. 
The field-free Hamiltonian follows $\hat{H_{0}} |\Psi^{N+1}_{0} \rangle = E_0 |\Psi^{N+1}_{0} \rangle$.
In general, $\hat{H}(t)$ can be partitioned as $\hat{H}(t) = \hat{H}_N(t) + \hat{H}_{F}(t) + \hat{V}_{Ne}$, where $\hat{H}_N(t)$ and $\hat{H}_{F}(t)$ acts only on the residual ion and the active electron, respectively,
and $\hat{V}_{Ne}$ is the interaction term between the residual ion and the active electron.
From the perspective of the active electron, this term represents the Coulomb potential and multipole potentials from the residual ion.
From the perspective of the bound electrons, this term represents electron correlation and could mix the ionic states.
The SFA in the single active electron picture implies that $\hat{V}_{Ne}$ is negligible for the active electron.
Extending that assumption to the multielectron picture, it means that $\hat{V}_{Ne}$ is also negligible for the bound electrons, and $\hat{H}(t)$ is separable.
Therefore, in Eq.~\eqref{eq:ME_amp}, we have 
\begin{align}
& \langle i | \otimes \langle \mathbf{p} + \mathbf{A}(t) | e^{-i \int_{t'}^{t} \hat{H}(t'')dt''} \nonumber \\
 &\approx  \langle i | e^{-i \int_{t'}^{t} \hat{H}_N(t'')dt''} 
\otimes   \langle \mathbf{p} + \mathbf{A}(t) | e^{-i \int_{t'}^{t} \hat{H}_{F}(t'')dt''}.
\label{eq:wrong}
\end{align}

\subsection{Dynamics during the tunneling~\label{sec3.1}}
Let us consider the dynamics during tunneling, so we are looking for the amplitude $a_{i \mathbf{p}}$ at the birth time $t_{i \mathbf{p}} = \mathrm{Re}(t')$ of an active electron with canonical momentum $\mathbf{p}$ and an ionic state $i$.
The instantaneous tunneling picture implies that the time lapse $\Delta t$ is imaginary during the tunneling process.
This corresponds to the complex position and momentum of the active electron under the barrier.
Meanwhile, the bound electrons do not see the barrier, and they should be frozen, experiencing a constant laser field at $t_{i \mathbf{p}}$.
Consequently, the ionic states right after tunneling should be the hole state of the residual ion.
We note that there are different representations of ionic states in the literature for multielectron tunneling ionization~\cite{Spanner2009, Torlina2012, Tolstikhin2014, Matsui2021}.
In general, $\hat{H}_N$ is not diagonal in the hole states basis in the presence of the laser field.
However, since $\mathrm{Re} (\Delta t) = 0$ during tunneling, there is no transition between hole states.
Therefore, we approximate $\hat{H}_N$ to be diagonal under the barrier, such that
\begin{align}
\langle i | e^{-i \int_{t'}^{t_{i \mathbf{p}}} \hat{H}_N(t'')dt''} 
\approx \langle i | e^{-i E_i (t_{i \mathbf{p}} - t')},
\end{align}
where $E_{i} = \langle i | \hat{H}_N(t_{i \mathbf{p}}) | i \rangle$ is the expectation value of the Hamiltonian $\hat{H}_N$ for hole state $i$.
Again, for simplicity, we assume that the laser field does not change the energy $E_{i}$.
Choosing $E_{0} = 0$, the ionization potential is simply $E_i$. 
Then, the ionization amplitude at $t_{i \mathbf{p}}$ becomes
\begin{align}
a_{i \mathbf{p}}(t_{i \mathbf{p}}) 
&= -i  \int_{t_i}^{t_{i \mathbf{p}}} dt' e^{-i \left\{  \int_{t'}^{t_{i \mathbf{p}}}  [ \frac{1}{2}[\mathbf{p} + \mathbf{A}(t'')]^{2} + E_{i}] dt''   \right\}} \nonumber \\
& \times c_{0}(t')  \langle i  | \otimes \langle \mathbf{p} + \mathbf{A}(t') | \hat{V}_{L}(t') |\Psi^{N+1}_{0} \rangle.
\label{eq:MESFA_amp}
\end{align}
Suppose $| i \rangle$ is the hole state formed by removing orbital $|\varphi_{i} \rangle$ from $|\Psi_{0} \rangle$.
The matrix element in Eq.~\eqref{eq:MESFA_amp} is then
\begin{align}
\langle i  | \otimes \langle \mathbf{p} + \mathbf{A}(t') | \hat{V}_{L}(t') |\Psi^{N+1}_{0} \rangle
= \langle \mathbf{p} + \mathbf{A}(t') | \hat{V}_{L}(t') |\varphi_{i} \rangle.
\end{align}
Therefore, Eq~\eqref{eq:MESFA_amp} is the same as Eq.~\eqref{eq:SAE_amp2}, except with the amplitude $c_{0}(t')$.
Following the treatment in Sec.~\ref{sec2}, the amplitude $a_{i \mathbf{p}}(t)$ is 
\begin{align}
a_{i \mathbf{p}}(t) &\approx \Theta(t-t_{i\mathbf{p}}) c_{0}(t_{i \mathbf{p}}) C_{i}(t_{i \mathbf{p}}) e^{-i S_{i \mathbf{p}}(t_{i \mathbf{p}})},
\label{eq:ME_a_amp}
\end{align}
with $E \to E_{i}$ in Eqs.~\eqref{eq:Re-S}, \eqref{eq:Im-S} and $S_{i \mathbf{p}}(t_{i \mathbf{p}}) = S_{i \mathbf{p}}(t_{i \mathbf{p}}, t_{i \mathbf{p}})$.
Again, the Heaviside step function ensures that the amplitude is zero before the birth time.
To account for the energy changes due to the laser field, $E_{i}$ can be taken as the time-dependent energy difference between the hole state and the neutral state.

\subsection{Dynamics after the tunneling~\label{sec3.2}}
Now we consider the bound electrons dynamics after the tunneling at $t_{i \mathbf{p}}$.
The total wave function is
\begin{align}
| \Psi(t) \rangle =  c_0 (t) | \Psi^{N+1}_0 \rangle + \int d\mathbf{p} \sum_i | \Psi_{i \mathbf{p}}(t) \rangle,
\label{eq:full-wavefunc}
\end{align}
with 
\begin{align}
| \Psi_{i \mathbf{p}}(t_{i \mathbf{p}})  \rangle =  a_{i \mathbf{p}} (t_{i \mathbf{p}}) |i \rangle \otimes | \mathbf{p + A}(t_{i \mathbf{p}}) \rangle
\end{align}
being the wave function right after forming the ionic hole state $i$ with an ionized electron of canonical momentum $\mathbf{p}$ and $c^{2}_{0}(t)$ being the population remaining in the neutral state.
For simplicity, we focus on the evolution of $| \Psi_{i \mathbf{p}}(t_{i \mathbf{p}})  \rangle$ and neglect the evolution of $| \Psi^{N+1}_0 \rangle$ for now, which leads to the birth of additional electrons.

The ionic hole state may not be an eigenstate of $\hat{H}_N(t)$, and the evolution of $|i \rangle$ can be represented as
\begin{align}
|\psi_{i} (t-t_{i \mathbf{p}}) \rangle = e^{-i \int_{t_{i \mathbf{p}}}^{t} \hat{H}_N(t'')dt''}  |i \rangle.
\label{eq:ion-evolve}
\end{align}
If the hole state is not an eigenstate of the field-free Hamiltonian, configuration interaction would lead to the mixing of eigenstates, similar to noble gas atoms with spin-orbit couplings~\cite{Pfeiffer2013}.
If the ionic states are dipole coupled, the evolution describes the postionization excitation dynamics.

Meanwhile, the evolution of the active electron is
\begin{align}
&e^{-i \int_{t_{i \mathbf{p}}}^{t} \hat{H}_{F}(t'')dt''} |  \mathbf{p + A}(t_{i \mathbf{p}}) \rangle \nonumber \\
&= e^{- \frac{i}{2} \int_{t_{i \mathbf{p}}}^{t}  [\mathbf{p} + \mathbf{A}(t'')]^{2} dt''}  
 | \mathbf{p + A}(t) \rangle.
\label{eq:el-evolve}
\end{align}

Therefore, the wave function $| \Psi_{i \mathbf{p}}(t) \rangle$ at a later time is
\begin{align}
| \Psi_{i \mathbf{p}}(t) \rangle &= a_{i \mathbf{p}} (t)   e^{- \frac{i}{2} \int_{t_{i \mathbf{p}}}^{t}  [\mathbf{p} + \mathbf{A}(t'')]^{2} dt''} \nonumber \\
&|\psi_{i}(t-t_{i \mathbf{p}}) \rangle  \otimes | \mathbf{p + A}(t) \rangle.
\label{eq:psi-ip}
\end{align}

The expression above is straightforward but laborious to solve: we need to calculate the birth time $t_{i \mathbf{p}}$ from Eqs.~\eqref{eq:gamma} and \eqref{eq:pz} and numerically evolve the ionic hole state from each birth time, but other terms are analytical.

\subsection{Overall tunneling dynamics~\label{sec3.3}}
Although we treated the dynamics during and after tunneling separately, these dynamics occur simultaneously.
Using Eqs.~\eqref{eq:ME_a_amp} and \eqref{eq:psi-ip}, the total wave function at time $t$ is
\begin{align}
| \Psi(t) \rangle &=  c_0 (t) | \Psi^{N+1}_0  \nonumber \rangle \\
& + \int d\mathbf{p} \sum_i a_{i \mathbf{p}} (t) e^{- \frac{i}{2} \int_{t_{i \mathbf{p}}}^{t}  [\mathbf{p} + \mathbf{A}(t'')]^{2} dt''}   \nonumber \\
&\times |\psi_{i}(t-t_{i \mathbf{p}}) \rangle  \otimes | \mathbf{p + A}(t) \rangle.
\label{eq:wavefunc-prop}
\end{align}

This expression is quite intuitive to understand.
For a hole state $i$ along with an ionized electron with canonical momentum $\mathbf{p}$,
\begin{enumerate}
    \item[(i)] if $t < t_{i \mathbf{p}}$, their amplitude is zero.
    \item[(ii)] if $t = t_{i \mathbf{p}}$ and $c_{0}(t_{i \mathbf{p}}) > 0$, the electron is ionized and the hole state is formed.
    \item[(iii)] if $t > t_{i \mathbf{p}}$, the ionized and bound electrons evolved under the laser field for a time $t - t_{i \mathbf{p}}$.
\end{enumerate}
The above occurs simultaneously with ionized electrons with canonical momentum $\mathbf{p}'$ from orbital $j$ at a different birth time $t_{j \mathbf{p}'}$.

To further simplify the theory, it is helpful to avoid propagating the ionic states for every possible birth time.
For neighboring orbitals contributing to ionization, their ionization potentials should be similar, and the birth times of an ionized electron with canonical momentum $\mathbf{p}$ should also be highly similar.
In the case of tunneling ionization of the $1\pi_{g}$ or $3\sigma_{u}$ orbitals of CO$_{2}$, it was shown that the birth time delay should be in tens of attoseconds~\cite{Shafir2012}.
In the next section, we examine the birth time delay for electrons from different orbitals.

\section{Zero Birth Delay Approximation~\label{sec4}}
In the derivation of the ionization amplitude, the use of the saddle point method leads to a longitudinal canonical momentum $p_{z}$ as a function of birth time $t$ and transverse canonical momentum $p_{\perp}$ \eqref{eq:pz}.
This implies that for a fixed $p_{\perp}$ and $t$, the spread of the longitudinal momentum is zero, leading to the Heaviside step function for the birth time $t_{\mathbf{p}}$ in Eq.~\eqref{eq:a_amp}.
However, the joint theory-experiment study by Pfeiffer \textit{et al.}~\cite{Pfeiffer2012} showed that its spread should be non-zero.
It is in accord with the energy-time uncertainty principle -- the birth of an electron with energy uncertainty 
\begin{align}
\sigma_{E} \sim E +  [\mathbf{p+A}(t)]^{2}/2
\end{align}
should be associated with an uncertainty in the birth time of 
\begin{align}
\sigma_{t} \sim 1 / \sigma_{E},
\end{align}
such that $p_{z}$ must be a distribution with a nonzero width.
Calculating the spread of $p_{z}$ is challenging; for example, see Refs.~\cite{Ivanov2005, Ivanov2024}.
However, if the birth delay between two electrons with a canonical momentum $\mathbf{p}$ from orbital $i$ and $j$ is similar to the birth time uncertainty,
\begin{align}
|t_{j \mathbf{p}} - t_{i \mathbf{p}}| / \sigma_{t}  \sim 1,
\end{align}
then it is reasonable to approximate $t_{j \mathbf{p}} \approx t_{i \mathbf{p}}$ within a certain time interval.
In the following, we consider ionization from two different orbitals, such that the ratio $e^{-2\kappa_{1}^{3}/3F_{0}} / e^{-2\kappa_{2}^{3}/3F_{0}} = 100$.
This leads to orbital 2 having a higher ionization potential, which contributes about 1\% to the total direct ionization.

\begin{figure}
\begin{center}
\includegraphics[width=8cm]{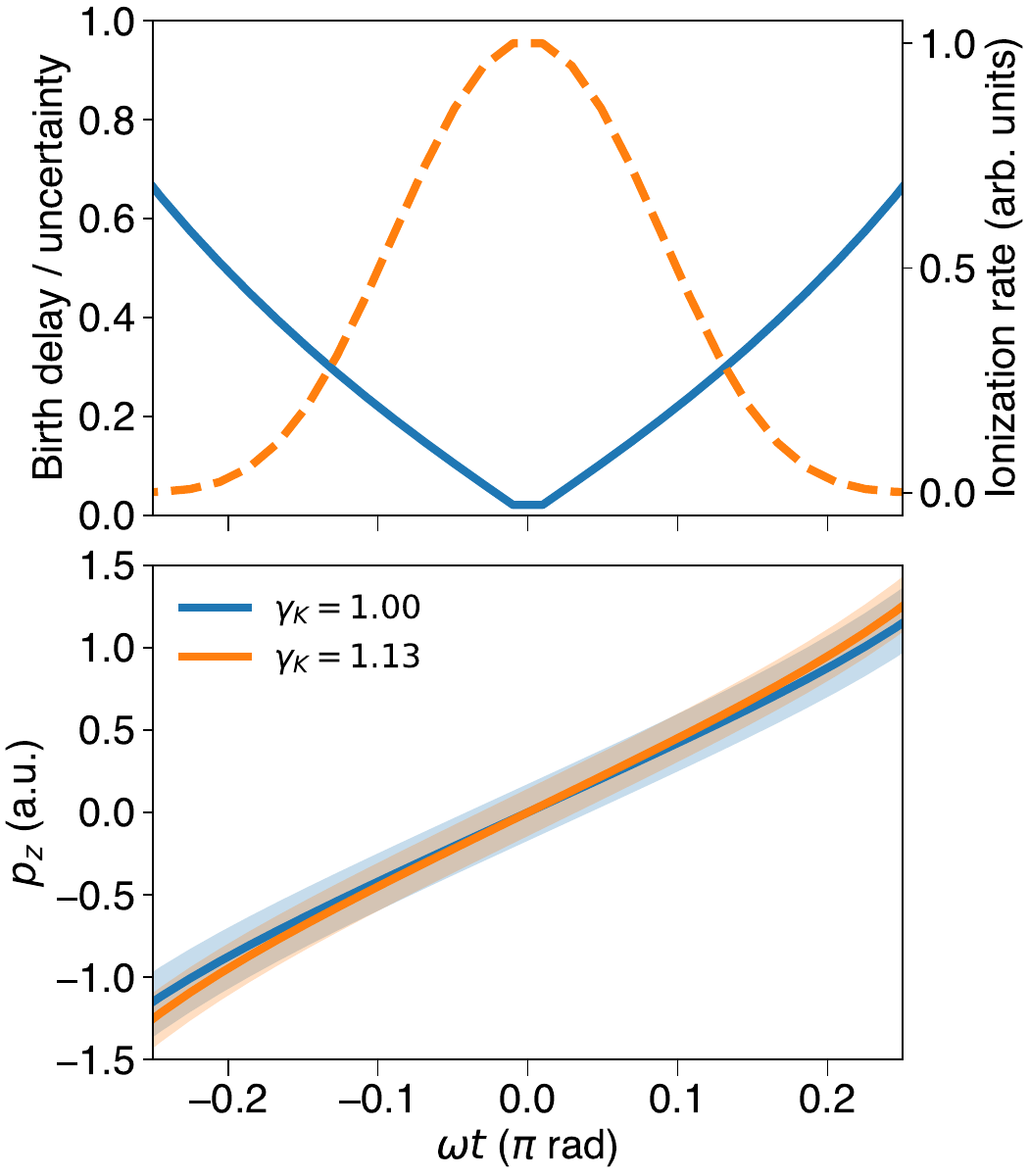}
\caption{Top: Ratio of the absolute value of birth delay to the birth time uncertainty of orbital 2 (solid line). The normalized ionization rate of orbital 2 (dashed line) is plotted to identify the relevant time interval.
Bottom: Longitudinal canonical momentum $p_{z}$ at $p_{\perp} = 0$ for orbitals 1 (blue line) and 2 (orange line), corresponding to Keldysh parameter $\gamma_{K}=$ 1.00 and 1.13.
The shaded area of the same color is the possible $p_{z}$ within the birth time uncertainty.
The laser wavelength is 800 nm with a peak intensity of $1 \times 10^{14}$ W/cm$^{2}$.}
\label{fig:pz}
\end{center}
\end{figure}

We now perform a numerical test under typical experimental conditions to determine whether the birth delay between orbitals 1 and 2 is significant.
Choosing the laser wavelength as 800 nm and peak intensity as $1 \times 10^{14}$ W/cm$^{2}$ and setting $\gamma_{K} = 1.00$ for orbital 1, we have $E_{1} = 12.0$ eV, $E_{2} = 15.3$ eV, and $\gamma_{K} = 1.13$ for orbital 2.
Taking the transverse canonical momentum $p_{\perp} = 0$ and using Eq.~\eqref{eq:pz}, we compute the longitudinal canonical momentum $p_{z}$ at different times.
To determine the relevant ionization time interval, the top panel of Fig.~\ref{fig:pz} shows the normalized ionization rate of orbital 2 via Eq.~\eqref{eq:NA-gamma}.
The ionization becomes negligible around $\omega t = \pm \pi/4$ rad.
The top panel of Fig.~\ref{fig:pz} also shows the ratio of the birth delay $|t_{2 p_{z}} - t_{1 p_{z}}|$ to the birth time uncertainty of the electron from orbital 2.
The ratio of the magnitude of birth delay to the birth time uncertainty is less than 1 in this interval, suggesting that birth delay is negligible.
On the other hand, we can analyze the change in $p_{z}$ by shifting the birth time based on the uncertainty from both orbitals 1 and 2.
In the bottom panel of Fig.~\ref{fig:pz}, the shaded regions show the possible $p_{z}$ values for orbitals 1 and 2 within the uncertainty of their birth times.
The two regions overlap significantly, suggesting that the birth delay is indeed negligible.

For orbitals of even higher ionization potential, their contribution to ionization would be negligible compared to the ionization rate to the ground state.
In the quasistatic limit, $\gamma_{K} \ll 1$, $p_{z}$ is identical for the orbitals, and the birth time delay goes to zero.
As a result, we approximate $t_{i \mathbf{p}} \approx t_{j \mathbf{p}}$ for any pair of orbitals, drop the channel subscript in the birth time, and represent $p_{z}$ for all orbitals with the $p_{z}$ from the highest occupied orbital hereafter.

\begin{figure}
\begin{center}
\includegraphics[width=8cm]{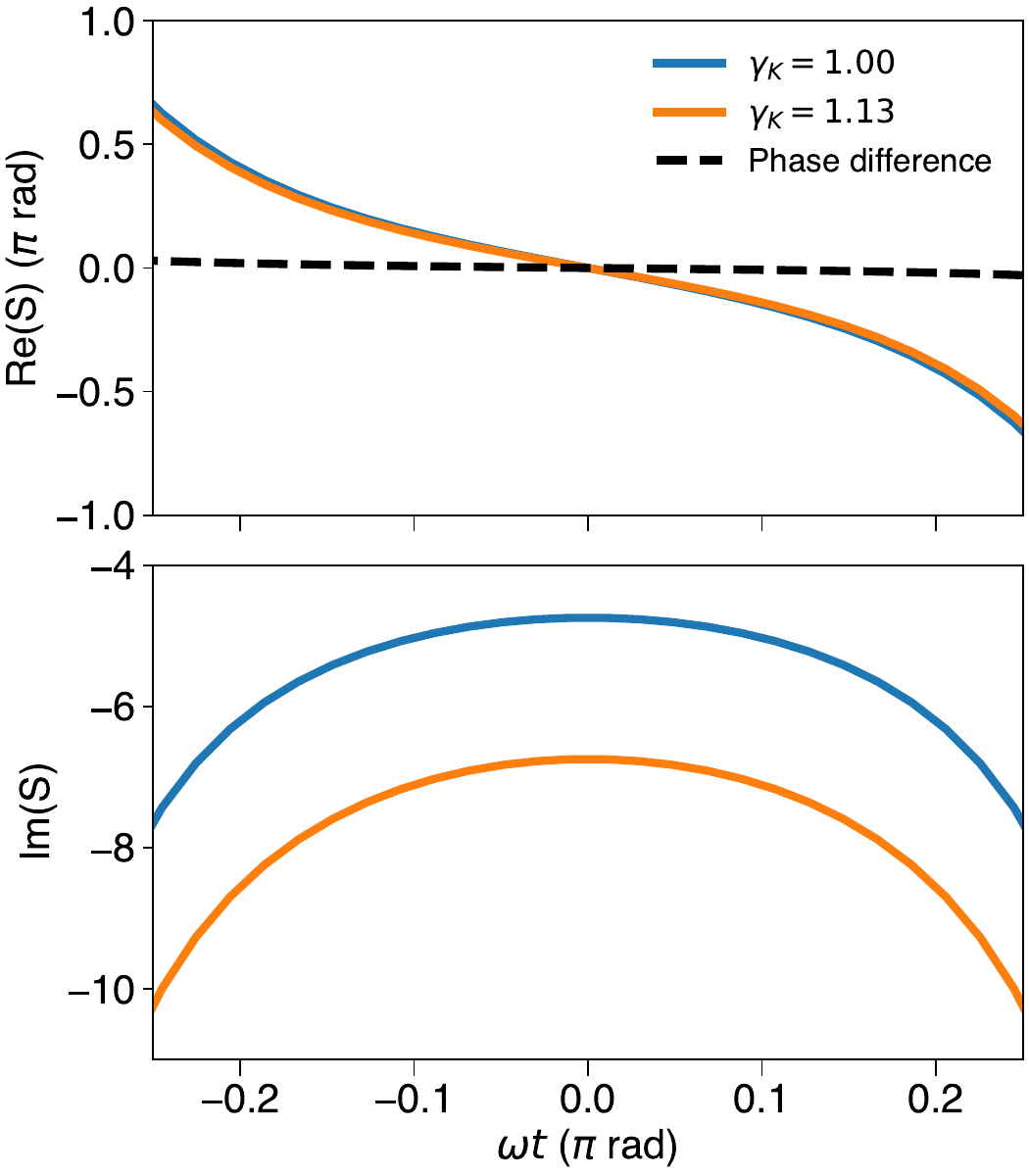}
\caption{Top: Under the barrier phase acquired by the active electron with $p_{\perp} = 0$ from orbitals 1 (blue line) and 2 (orange line), corresponding to Keldysh parameter $\gamma_{K}=$ 1.00 and 1.13.
The longitudinal canonical momentum is approximated to be identical for the two orbitals.
The dashed line is the phase difference Re$(S_{1} - S_{2})$.
Bottom: Imaginary part of the action with $p_{\perp} = 0$ from orbitals 1 (blue line) and 2 (orange line).
The laser parameters are identical to Fig.~\ref{fig:pz}.}
\label{fig:S-diff}
\end{center}
\end{figure}

The zero birth delay approximation has important consequences for the tunneling and Coulomb phases of the ionization amplitude.
The approximation implies that $\sqrt{1 + \gamma^{2}_{1, \perp t}} \approx \sqrt{1 + \gamma^{2}_{2, \perp t}}$ [c.f. Eqs.~\eqref{eq:gamma}, \eqref{eq:pz}].
Therefore, according to Eq.~\eqref{eq:Re-S}, the real part of the under the barrier action of the active electron from orbitals 1 and 2 should be nearly identical.
The top panel of Fig.~\ref{fig:S-diff} displays the real part of the under the barrier action of the active electron with $p_{\perp} = 0$ from orbitals 1 and 2, calculated by using the same $p_{z}$.
We observe that the phases of the active electron from orbitals 1 and 2 have negligible differences, as expected.
On the other hand, the imaginary part of the action depends on $\gamma_{\perp t}$, such that even using the same $p_{z}$, it should be different between orbitals 1 and 2.
The bottom panel of Fig.~\ref{fig:S-diff} shows that it is indeed the case at $p_{\perp} = 0$, where the magnitude of Im$(S_{2})$ is 42.2\% larger than Im$(S_{1})$ at the peak field. 
In addition, the zero birth delay approximation implies that the sub-barrier trajectory of the active electron from orbitals 1 or 2 is identical.
Consequently, their Coulomb correction in the action is identical~\cite{Yan2012, Popruzhenko2014}.
Indeed, the magnitude of the Coulomb correction factors in the ionization amplitude, $( 2\kappa^{3}/F_{0} )^{Z/\kappa} ( 1 + 2 e^{-1} \gamma_{K})^{-Z/\kappa}$,  differ by only around 6\% between orbitals 1 and 2.
The phases from the Coulomb correction in the quasistatic limit, $\pi Z /\kappa$~\cite{Tolstikhin2011, Yan2012}, differ by about 12\% between the two orbitals. 

\section{Density matrix approach for multielectron tunneling ionization~\label{sec5}}
With the use of zero birth delay approximation, the birth time becomes independent of ionic states, and Eq.~\eqref{eq:wavefunc-prop} can be rewritten as
\begin{align}
| \Psi(t) \rangle &=  c_0 (t) | \Psi^{N+1}_0 \rangle  \nonumber \\
& + \int d\mathbf{p}  \sum_i   a_{i \mathbf{p}} (t) e^{- \frac{i}{2} \int_{t_{\mathbf{p}}}^{t}  [\mathbf{p} + \mathbf{A}(t'')]^{2} dt''}  \nonumber \\
&\times  |\psi_{i}(t-t_{\mathbf{p}}) \rangle  \otimes  | \mathbf{p + A}(t) \rangle.
\label{eq:wavefunc-prop2}
\end{align}
The above expression has been used either implicitly or explicitly over the years.
For example, in multichannel high-harmonic spectroscopy~\cite{Smirnova2009, Kraus2015, He2022, He2023}, the zero birth delay approximation was used to map the birth time and rescattering energy of ionized electron from different orbitals, and the residual ion was evolved as in Eq.~\eqref{eq:ion-evolve}.
Eq.~\eqref{eq:wavefunc-prop2} in the quasistatic limit was also used without explicitly describing the ionized electron in studies of the positionization excitation of N$_2^{+}$ in Refs.~\cite{Xu2015, Zhang2017, Xu2022}.

\subsection{Reduced ionic density matrix~\label{sec5.1}}
To focus on the dynamics of the bound electrons, we can simplify Eq.~\eqref{eq:wavefunc-prop2} by tracing out the ionized electron $ | \mathbf{p + A}(t) \rangle$ and investigate the dynamics of the reduced ionic density matrix  $\hat{\rho}_{N}$,
\begin{align}
\hat{\rho}_{N}(t) &\equiv \int d\mathbf{p} \, \langle \mathbf{p+A}(t)| \Psi(t) \rangle  \langle \Psi(t)  | \mathbf{p+A}(t) \rangle \\
& = \int d\mathbf{p} \, \sum_{i,j}   a_{i \mathbf{p}}(t) a^{\ast}_{j \mathbf{p}} (t) \nonumber \\
& \times |\psi_{i}(t-t_{\mathbf{p}}) \rangle \langle \psi_{j}(t-t_{\mathbf{p}}) |.
\end{align}

From Eqs.~\eqref{eq:ME_a_amp} and \eqref{eq:ion-evolve}, we have
\begin{align}
\frac{\partial}{\partial t} a_{i \mathbf{p}}(t) = \delta(t - t_{\mathbf{p}})  c_0 (t_{\mathbf{p}})  C_{i}(t_{\mathbf{p}}) e^{-i S_{i \mathbf{p}}(t_{\mathbf{p}})}
\end{align}
and
\begin{align}
\frac{\partial }{\partial t}|\psi_{i}(t-t_{\mathbf{p}}) \rangle = -i \hat{H}_{N}(t) |\psi_{i}(t-t_{\mathbf{p}}) \rangle.
\end{align}

Therefore, the evolution of $\hat{\rho}_{N}$ is 
\begin{align}
\frac{\partial}{\partial t} \hat{\rho}_{N}(t) = -i [\hat{H}_{N}(t), \hat{\rho}_{N}(t)] + \hat{\Gamma}(t),
\label{eq:EOM}
\end{align}
where 
\begin{align}
\Gamma_{ij}(t) &=  \int d\mathbf{p} \, c^{2}_0 (t_{\mathbf{p}}) \sum_{i,j} 2 C_{i}(t_{\mathbf{p}}) C^{\ast}_{j}(t_{\mathbf{p}}) \nonumber \\
&\times  \delta(t - t_{\mathbf{p}}) \Theta(t - t_{\mathbf{p}}) e^{-i \left[S_{i \mathbf{p}}(t_{\mathbf{p}}) - S^{\ast}_{j \mathbf{p}}(t_{\mathbf{p}})\right]}
\end{align}
is the ionization matrix.
In the above, the exponent can be approximated as $\mathrm{Im} \left[S_{i \mathbf{p}}(t_{\mathbf{p}}) + S_{j \mathbf{p}}(t_{\mathbf{p}})\right]$ because the tunneling phase between orbitals $i$ and $j$ are approximately equal under the zero birth delay approximation.
We can then apply the saddle point method to the $p_{\perp}$ integral and transform the $p_{z}$ integral to a birth time integral as in Eqs.~\eqref{eq:p-perp-saddle} and \eqref{eq:b4-lastsaddle}, respectively. 

In Appendix~\ref{appendix2}, we showed that, since the Coulomb phase between orbitals $i$ and $j$ also cancels out, the ionization matrix $\Gamma_{ij}$ is given by
\begin{align}
\Gamma_{ij}(t) = \rho_{0}(t) \frac{B_{i} B^{\ast}_{j}}{|B_{i}| |B_{j}|} \sqrt{w_{i}(t) w_{j}(t)},
\label{eq:coherence-rate}
\end{align}
where the neutral population is
\begin{align}
\rho_{0}(t) = c^{2}_{0}(t) = \exp\left[-\int_{t_{i}}^{t} dt' \sum_i w_{i}(t') \right].
\label{eq:pop0}
\end{align}
The subscript $i$ or $j$ (except for $t_{i}$) means $E$, $\kappa$, and $\gamma_{K}$ should be replaced by their respective values for orbital $i$ or $j$ in Eqs. \eqref{eq:Bm} and \eqref{eq:NA-gamma}.
Since the diagonal and off-diagonal elements of the density matrix are the population and coherence in the system, the diagonal elements of $\Gamma$ represent the ionization rate for each orbital, while the off-diagonal elements represent the ionization coherence buildup rate.

As a result, we demonstrated that directly evolving the reduced ionic density matrix is \textit{equivalent} to evolving the full wave function and then tracing out the ionized electron.
In other words, these two equations are direct consequences of the multielectron SFA and the zero birth delay approximation.
In the quasistatic limit $\gamma_{K} \ll 1$, Eq.~\eqref{eq:coherence-rate} reduces to the ionization matrix in Refs.~\cite{Yuen2023b, Yuen2024b} with $m = 0$, confirming the validity of previous density matrix approaches based on ADK rates.

 Eqs.~\eqref{eq:EOM} and \eqref{eq:coherence-rate} will be referred to as the density matrix approach for strong field ionization, or the DM-SFI theory.
Compared with the wave function approach, the DM-SFI theory is more convenient to use because it does not require explicit tracking of the ionized electron, and the evolution of the bound electrons can be performed only once, rather than for each birth time.
The physical meaning of the equation of motion is as follows: At each time $t$, the neutral state would be tunnel ionized, injecting population and coherence for the reduced ionic density matrix, and the existing ionic population and coherence would evolve under the von Neumann-Liouville equation.

Beyond confirming the validity of density matrix approaches for subcycle ionic dynamics, this work improves on previous density matrix models.
Compared to Refs.~\cite{Pfeiffer2013, Pabst2016}, the current model accounts for complex phase factors for polar molecules and the effects of dipole couplings.
It includes tunnel ionization coherence $\Gamma_{ij}$, which was neglected in Refs.~\cite{Zhang2020, Tikhonchuk2021, Xu2022, Lei2022, Yuen2022, Yuen2023, Yuen2024a, Chen2024, Jia2024, Zhao2025, Jia2025}.
For models that include tunnel ionization coherence~\cite{Yuen2023b, Yuen2024b, Yuen2024c, Zhou2025, Li2025, Xue2025}, the current model improves the accuracy by using the nonadiabatic ionization rate~\eqref{eq:NA-gamma}.

\subsection{Reduced electron density matrix~\label{sec5.2}}
Recent interests in measuring the quantum state of photoelectrons~\cite{Laurell2025} prompt the question of whether direct electrons from tunneling ionization carry information about the electronic coherence in the residual ion.
Previous photoelectron angular distributions measurements~\cite{Liu2012, Li2015} demonstrated that direct electrons carry no structural information about the targets.
For completeness in the analysis of the density matrix, we investigate whether the reduced electron density matrix $\hat{\rho}_{e}$ encodes the electronic coherence of the residual ion.

The reduced electron density matrix $\hat{\rho}_{e}$ is defined as
\begin{align}
\hat{\rho}_{e}(t) \equiv \sum_{j} \langle j | \Psi(t) \rangle \langle \Psi(t) | j \rangle.
\end{align}
Since the action after the tunnel exit and the Volkov wave function is independent of the ionic states, we contract the sum over $j$ using $\sum_{j} | j \rangle  \langle j|  = 1$ and have
 \begin{align}
\hat{\rho}_{e}(t) &= \int \int d\mathbf{p} \,d\mathbf{p}'\,    e^{- \frac{i}{2} \int_{t_{\mathbf{p}}}^{t}  [\mathbf{p} + \mathbf{A}(t'')]^{2} dt''} \nonumber \\ 
&\times  e^{ \frac{i}{2} \int_{t_{\mathbf{p'}}}^{t}  [\mathbf{p}' + \mathbf{A}(t'')]^{2} dt''} | \mathbf{p + A}(t) \rangle \langle  \mathbf{p' + A}(t) | \nonumber \\
&\times \sum_{i,i'} a_{i \mathbf{p}} (t) a^{\ast}_{i' \mathbf{p'}} (t)  \langle \psi_{i'}(t-t_{\mathbf{p}'}) |\psi_{i}(t-t_{\mathbf{p}}) \rangle.
\label{eq:el-dm} 
\end{align}

We first analyze the population of ionized electrons with canonical momentum $\mathbf{p}$.
With the same $\mathbf{p}$, the phase from the action cancels out.
We also have
\begin{align}
\langle \psi_{i'}(t-t_{\mathbf{p}}) |\psi_{i}(t-t_{\mathbf{p}}) \rangle = \delta_{i i'}
\end{align}
since the evolution operator cancels out.
Therefore, the population is simply
\begin{align}
\rho_{e, \mathbf{p} \mathbf{p}}(t) = \sum_{i} |a_{i \mathbf{p}} (t)|^{2}, 
\end{align}
which is the total ionization probability for electrons with canonical momentum $\mathbf{p}$ if $t > t_{\mathbf{p}}$.
It has been known that the photoelectron momentum distribution largely depends on the Keldysh parameter $\gamma_{K}$~\cite{Liu2012, Li2015}.
Since different contributing orbitals have similar $\gamma_{K}$ values, unless the above-threshold ionization peaks are well-resolved~\cite{Boguslavskiy2012}, it is challenging to probe the electronic structure of the residual ion using photoelectron momentum spectroscopy.

On the other hand, while structural information is contained in terms of $a_{i \mathbf{p}} (t) a^{\ast}_{i' \mathbf{p'}} (t)  \langle \psi_{i'}(t-t_{\mathbf{p}'}) |\psi_{i}(t-t_{\mathbf{p}}) \rangle$ in Eq.~\eqref{eq:el-dm}, such information maybe challenging to extract.
This is because the phases of the ionization amplitude and the dynamics phases of the ionized electron at different canonical momenta do not cancel out, resulting in rapidly oscillating functions.
Coulomb corrections to the ionization amplitude will be required to compare with experimental photoelectron momentum spectra~\cite{Popruzhenko2014} and will further complicate the extraction.
However, indirect structural information can be obtained from the purity of the reduced electron density matrix, $\mathrm{Tr}(\hat{\rho}_{e}^{2})$, which is less than one if more than one ionic state is populated after tunneling.

Therefore, unlike the photoelectrons created from an attosecond pulse train~\cite{Laurell2025}, we expect that it would be difficult to extract coherence information from strong field ionized electrons.

\section{Application of the DM-SFI theory to molecules~\label{sec6}}
Recent experimental evidence shows the creation of excited state populations and ionic coherence by a strong infrared laser pulse in N$_{2}^{+}$~\cite{Kleine2022, Chen2024, Gao2025, Danylo2025} and CO$_{2}^{+}$~\cite{Gao2025co2}.
In this section, we apply the DM-SFI theory to investigate the ionic dynamics in the multielectron tunneling of these molecules.
The nuclei are assumed to be frozen during the laser-molecule interaction.
Computational details for the DM-SFI theory and electronic structures of these molecules are provided in Appendix~\ref{appendix3}.

\subsection{N$_{2}^{+}$ molecule~\label{sec6.1}}
Excited state population and ionic coherence in N$_{2}^{+}$ created by an intense laser pulse have been studied extensively both theoretically and experimentally due to their relevance to air lasing~\cite{Yao2011, Liu2013, Liu2015, Xu2015, Yao2016, Zhang2020, Tikhonchuk2021, Xu2022, Kleine2022, Lei2022, Chen2024, Li2025, Zhou2025, Gao2025, Danylo2025}.
The mechanisms behind the enhanced excited state population~\cite{Xu2015, Zhang2020, Yuen2023b, Li2025}, wavelength dependence enhancement of lasing~\cite{Chen2024, Zhou2025}, as well as the CEP dependence of the ionic coherence~\cite{Gao2025} have been well understood using a density matrix approach with ADK rates.
The key question here is whether the qualitative behavior of subcycle ionic dynamics would be altered if the nonadiabatic rates are used.
To answer that question, it is sufficient to investigate only the subcycle dynamics in N$_{2}^{+}$ at a particular orientation with respect to the laser polarization.
The laser pulse considered here is a 3.7 fs FWHM Gaussian pulse with a central wavelength of 900 nm, a peak intensity of $2 \times 10^{14}$ W/cm$^{2}$, and a zero carrier-envelope phase.

\begin{table}
  \centering 
  \caption{The electronic states of N$_{2}^{+}$ considered in this work. The Keldysh parameter $\gamma_{K}$ and $F_{0}/\kappa^{3}$ correspond to a 900 nm laser field with a peak intensity of $2 \times 10^{14}$ W/cm$^{2}$.}
  \begin{tabular}{ccccccc}
\hline
\hline
State  &  Abbrev. & $E$ (eV) & Orbital & $\gamma_{K}$ & $F_{0}/\kappa^{3}$ \\
$X^{2}\Sigma_{g}^{+}$ & $X$  & 15.6 & $3\sigma_{g}$ & 0.717 & 0.0616 \\
$A^{2}\Pi_{u\pm}$ & $A_{\pm}$  & 16.9 & $1\pi_{u\pm}$ & 0.748 & 0.0544 \\
$B^{2}\Sigma_{u}^{+}$  & $B$ & 18.8 & $2\sigma_{u}$ & 0.787 & 0.0467 \\
\hline
\end{tabular}
\label{tab:N2}
\end{table}

Table~\ref{tab:N2} summarizes the electronic structures of N$_{2}^{+}$.
The highest occupied molecular orbital (HOMO) of N$_{2}$ is $3\sigma_{g}$, which has a binding energy of 15.6 eV.
The HOMO-1 ($1\pi_{u\pm}$) and HOMO-2 ($2\sigma_{u}$) have a binding energy of 16.9 and 18.8 eV, respectively.
Tunneling ionization of the HOMO, HOMO-1, and HOMO-2 lead to the $X^{2}\Sigma_{g}^{+}$, $A^{2}\Pi_{u\pm}$, and $B^{2}\Sigma_{u}^{+}$ states of N$_{2}^{+}$.
The parameters $F_{0}/\kappa_{i}^{3}$ are less than 0.1, but the Keldysh parameters at the peak field are around 0.7 to 0.8.
Therefore, it is valid to employ the nonadiabatic rate \eqref{eq:NA-gamma}, but the ADK rate may be inaccurate.

Furthermore, the $X$ and $A_{\pm}$ states are dipole coupled with $\textbf{d} = \mp 0.177 \, \mathbf{\hat{x}}$~\cite{Langhoff1987, Yuen2023b}, while the $X$ and $B$ states are dipole coupled with $\textbf{d} = 0.75 \,\mathbf{\hat{z}}$~\cite{Langhoff1988}.
Therefore, $A$ and $B$ states could be formed by postionization excitation from the $X$ state.

\begin{figure}[t]
\begin{center}
\includegraphics[width=7.5cm]{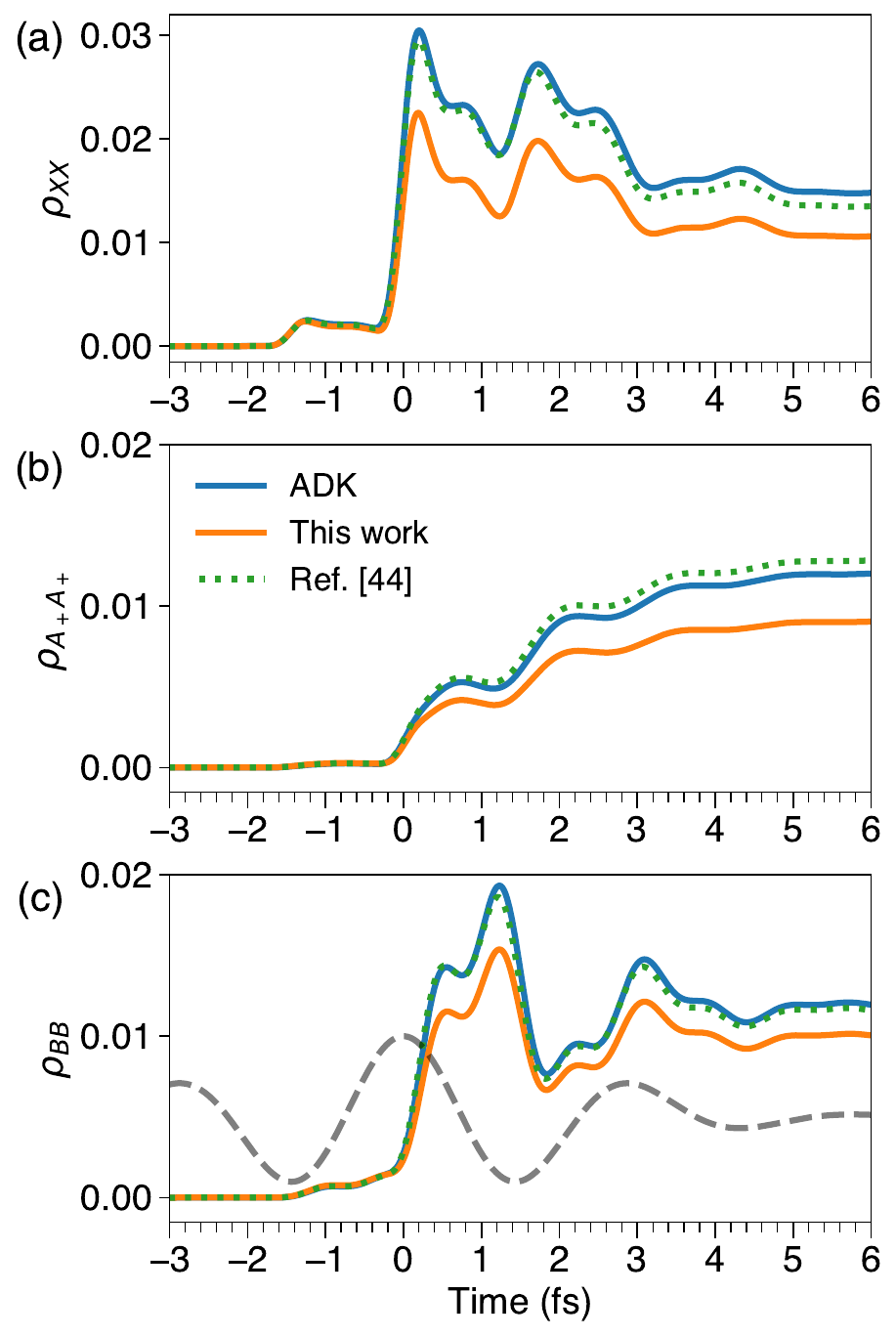}
\caption{Subcycle dynamics of the population of (a) $X^{2}\Sigma_{g}^{+}$, (b) $A^{2}\Pi_{u+}$, and (c) $B^{2}\Sigma_{u}^{+}$ state of N$_{2}^{+}$ at 45 degrees, calculated using the DM-SFI theory with ADK (blue lines) and nonadiabatic rates~\eqref{eq:NA-gamma} (orange lines) and Eqs. (5) and (8) in Ref.~\cite{Yuen2024b} (green dotted lines). The dashed line depicts the 3.7 fs FWHM Gaussian pulse at 900 nm, with a peak intensity of $2 \times 10^{14}$ W/cm$^{2}$ and a zero carrier-envelope phase.}
\label{fig:N2-subcycle}
\end{center}
\end{figure}

Figure~\ref{fig:N2-subcycle} displays the population of the $X$, $A_{+}$, and $B$ state of N$_{2}^{+}$ aligned at 45 degrees with the laser polarization. 
For the $X$, $A$, and $B$ states, the final population obtained from the ADK rates are 28.3, 24.8, and 15.8\% higher than those from the nonadiabatic rates.
The transient populations from the ADK rates are consistently larger than those from the nonadiabatic rates.
As a result, the ADK rates are not quantitatively accurate with the considered laser parameters.

Regarding the subcycle dynamics, in Fig.~\ref{fig:N2-subcycle}, population bursts due to tunneling occur at around $t=-1.4, 0,$ and 1.4 fs for all three states. 
At $t=0.6$ fs, the drop of the $X$ state population corresponds to the increase of populations in 
both the $A_{+}$ and $B$ states.
On the other hand, between $t=1.2$ to 1.8 fs, the sharp decrease of the $B$ state population corresponds to the increase of the $X$ and $A_{+}$ states population.
Therefore, the results reveal complex excitation dynamics between three ionic states.
The modulations in the populations obtained from the ADK and nonadiabatic rates are identical, indicating that the model with ADK rates accurately captures the qualitative behavior of the subcycle dynamics.
This can be readily understood from Eq.~\eqref{eq:coherence-rate}: The tunnel ionization coherence from both rates has the same phase, leading to the same subcycle dynamics.

Additionally, the phase in Eq.~\eqref{eq:coherence-rate} differs from Eqs. (5) and (8) in Ref.~\cite{Yuen2024b}, where the Coulomb phase $\pi Z/\kappa_{i}$~\cite{Tolstikhin2011, Yan2012} is included for adiabatic tunneling.
In Fig.~\ref{fig:N2-subcycle}, we also compare the populations for the $X$, $A_{+}$, and $B$ states calculated using Eq.~\eqref{eq:coherence-rate} with the ADK rates and those calculated using Eqs. (5) and (8) in Ref.~\cite{Yuen2024b}.
The subcycle dynamics are clearly identical.
We found that the final populations differ by only 8.96, 6.77, and 2.70\%, respectively.
These observations align with the conclusion in Sec.~\ref{sec4}, where the zero birth delay approximation implies that the difference of the Coulomb phase between active electrons from different orbitals is negligible.

\begin{figure*}
\begin{center}
\includegraphics[width=13.cm]{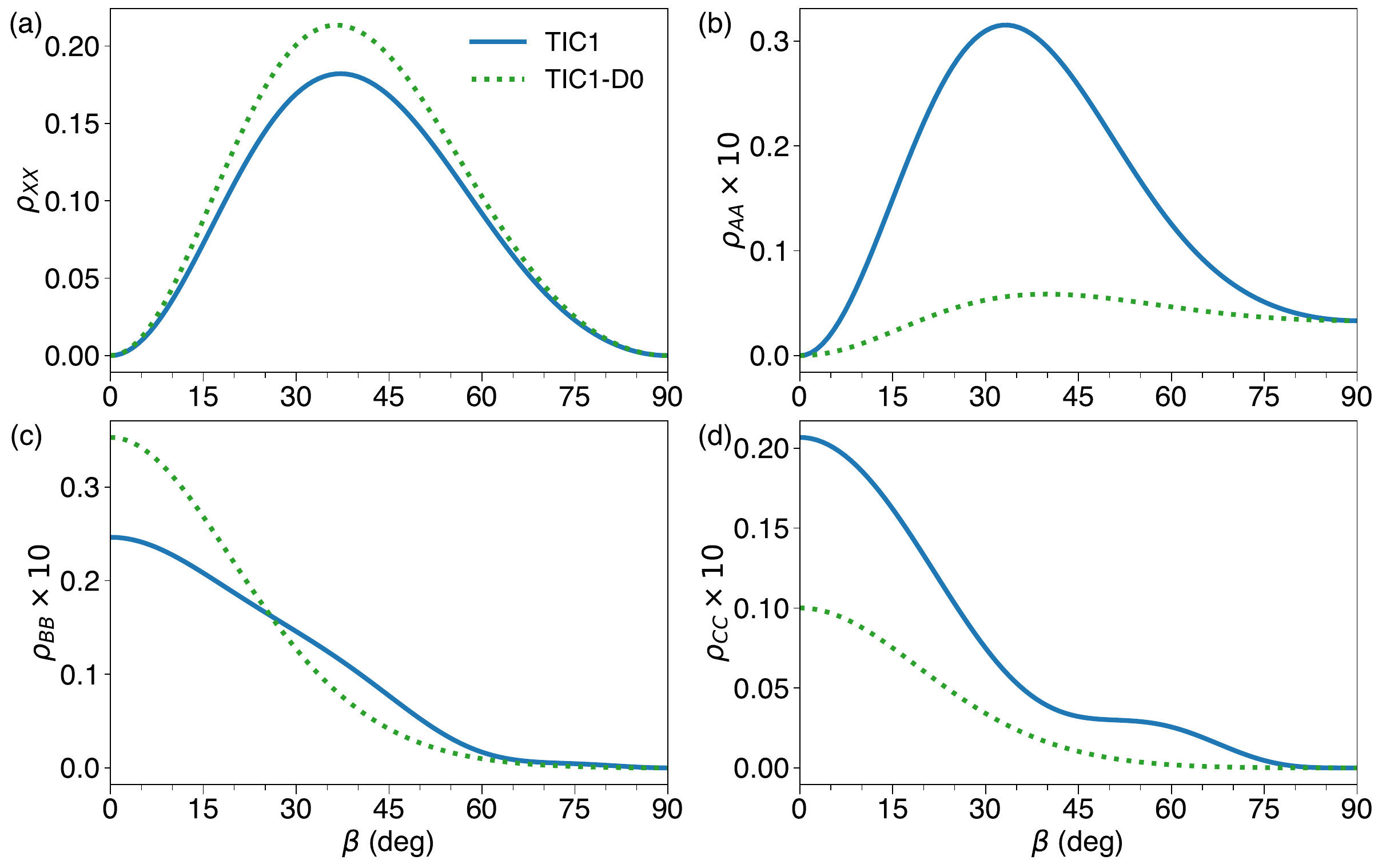}
\caption{Alignment dependence of the final population of (a) $X^{2}\Pi_{g}$, (b) $A^{2}\Pi_{u}$, (c) $B^{2}\Sigma_{u}^{+}$, and (d) $C^{2}\Sigma_{g}^{+}$ state of CO$_{2}^{+}$, calculated using the DM-SFI theory with nonadiabatic rates~\eqref{eq:NA-gamma}.
The excited state formation mechanisms are illustrated by considering the full theory (TIC1, solid lines) and the theory with tunnel ionization coherence but without dipole couplings (TIC1-D0, dotted lines). The laser parameters are identical to Fig.~\ref{fig:N2-subcycle}.}
\label{fig:CO2-ang-dist}
\end{center}
\end{figure*}

The above results suggest that the DM-SFI theory with ADK rates is qualitatively accurate.
This explains why similar theories have been successful in interpreting experimental observations~\cite{Xu2015, Zhang2020, Li2025, Chen2024, Zhou2025, Gao2025}.
A significant advantage of using the ADK rates is that they allow the use of laser fields with arbitrary pulse shapes.
As demonstrated in a recent study by Weckwerth \textit{et al.}~\cite{Weckwerth2025}, this theory can be applied to retrieved laser pulses to reproduce qualitative features observed in experimental measurements.
This contrasts with the use of the nonadiabatic rate, in which the laser field is restricted to a pulse envelope with a sinusoidal carrier.
However, if quantitative accuracy is required, such as predicting the onset of population inversion between the $B$ and $X$ state, we recommend using the nonadiabatic rates.

\subsection{CO$_{2}^{+}$ molecule~\label{sec6.2}}
The multielectron tunneling dynamics of the CO$_{2}$ molecule have been studied in the context of high-harmonic spectroscopy~\cite{Smirnova2009, Ruberti2018, He2022, Shu2022}.
The alignment dependence of the tunneling ionization rate of CO$_{2}$ has also been investigated~\cite{Pavicic2007, Spanner2009, Zhao2010, Lam2020, Zhu2024}.
It has been known that the $X^{2}\Pi_{g \pm}$, $A^{2}\Pi_{u \pm}$, $B^{2}\Sigma_{u}^{+}$, and $C^{2}\Sigma_{g}^{+}$ of CO$_{2}^{+}$ could be formed upon strong field ionization~\cite{Ruberti2018, Shu2022}.
The relative phase of the ionic wave function was retrieved from the experiment, and subcycle electron movies in CO$_2^+$ were created~\cite{Smirnova2009, He2022, Shu2022}.
Recently, the lasing experiment by Gao \textit{et al.}~\cite{Gao2025co2} demonstrated that multielectron tunneling dynamics could create ionic coherence between the $X^{2}\Pi_{g \pm}$ and $A^{2}\Pi_{u \pm}$ states.
Therefore, it is of interest to apply the DM-SFI theory to CO$_2$ to explore the mechanism behind the excited state population and the ionic coherence buildup.
The laser parameters considered are identical to the N$_{2}$ case.

\begin{table}
  \centering 
  \caption{The electronic states of CO$_{2}^{+}$ considered in this work. The ionization potentials $E$ are obtained from Ref.~\cite{Turner1970}. The Keldysh parameter $\gamma_{K}$ and $F_{0}/\kappa^{3}$ correspond to a 900 nm laser field with a peak intensity of $2 \times 10^{14}$ W/cm$^{2}$.}
  \begin{tabular}{ccccccc}
\hline
\hline
State  &  Abbrev. & $E$ (eV) & Orbital & $\gamma_{K}$ & $F_{0}/\kappa^{3}$ \\
$X^{2}\Pi_{g\pm}$ & $X_{\pm}$  & 13.8 & $1\pi_{g\pm}$ & 0.675 & 0.0739 \\
$A^{2}\Pi_{u\pm}$ & $A_{\pm}$  & 17.7 & $1\pi_{u\pm}$ & 0.765 & 0.0509 \\
$B^{2}\Sigma_{u}^{+}$  & $B$ & 18.2 & $3\sigma_{u}$ & 0.776 & 0.0488 \\
$C^{2}\Sigma_{g}^{+}$  & $C$ & 19.4 & $4\sigma_{g}$ & 0.800 & 0.0443 \\
\hline
\end{tabular}
\label{tab:CO2}
\end{table}

Table~\ref{tab:CO2} summarizes the electronic structures of CO$_{2}^{+}$ states considered here. 
The HOMO ($1\pi_{g}$), HOMO-1 ($1\pi_{u}$), HOMO-2 ($3\sigma_{u}$), and HOMO-3 ($4\sigma_{g}$) of CO$_{2}$ have ionization potentials of 13.8, 17.7, 18.2, and 19.4 eV, respectively~\cite{Turner1970}.
Tunneling ionization of the HOMO, HOMO-1, HOMO-2, and HOMO-3 leads to the formation of the $X^{2}\Pi_{g\pm}$, $A^{2}\Pi_{u\pm}$, $B^{2}\Sigma_{u}^{+}$, and $C^{2}\Sigma_{g}^{+}$ states of CO$_{2}^{+}$.
The parameters $F_{0}/\kappa_{i}^{3}$ are less than 0.1, but the Keldysh parameters could be as large as 0.8.
Therefore, the nonadiabatic rate is valid, but the ADK rate may again be inaccurate.

After tunneling, the $X_{\pm}A_{\pm}$, $X_{\pm}B$, and  $BC$ states are coupled by dipole moments $\textbf{d} = 0.534 \mathbf{\hat{z}}$, $\mp 0.16 \mathbf{\hat{x}}$, and $0.936 \mathbf{\hat{z}}$, respectively.
While the $A_{\pm}C$ states are also dipole coupled, the dipole moment is small ($\mp 6.15 \times 10^{-3} \mathbf{\hat{x}}$) and is neglected.

To understand the mechanisms of excited state formation and coherence buildup, we introduce three levels of theory.
The TIC1-D0 model includes tunnel ionization coherence but neglects dipole couplings by setting $\textbf{d} = 0$.
The TIC0 model considers the dipole couplings but neglects the tunnel ionization coherence, i.e., the off-diagonal elements of $\Gamma_{ij}$~\eqref{eq:coherence-rate}.
The complete theory is denoted as the TIC1 model, as described by Eqs.~\eqref{eq:EOM} and \eqref{eq:coherence-rate}.
In general, comparing excited state populations between the TIC1-D0 and TIC1 models indicates whether excited states originate from tunneling or positionization excitation.
Comparing the magnitudes of ionic coherences among the TIC1-D0, TIC0, and TIC1 models reveals the roles of tunneling and postionization excitation in coherence buildup.

\subsubsection{Excited state population~\label{sec6.2.1}}
Figure~\ref{fig:CO2-ang-dist} compares alignment dependence on the final population of $X$, $A$, $B$, and $C$ states of CO$_{2}^{+}$, calculated using the TIC1 and TIC1-D0 models with the nonadiabatic rates.
The population accounts for the two-fold degeneracy of the $\Pi$ states.
Each excited state population is about an order of magnitude smaller than that of the $X$ state.

$A$ state --- In the TIC1 model, the alignment dependence of the $A$ state population is similar to that of the $X$ state at angles less than about 60$^{\circ}$.
At larger angles, the population from the TIC1 model converges with that from the TIC1-D0 model.
The peak population of the $A$ state in the TIC1 model is 81.4\% higher than in the TIC1-D0 model.
These results indicate that the $A$ state population is primarily formed by excitation from the $X$ state.

$B$ state --- The $B$ state population is peaked at 0$^{\circ}$ in both TIC1-D0 and TIC1 models, so that its alignment dependence is distinct from that of the $X$ state.
At angles larger than 60$^{\circ}$, $X$ state population is suppressed, and the alignment dependence in the TIC1 model converges with that of the TIC1-D0 model.
Therefore, the enhancement at around 30$^{\circ}$ to 60$^{\circ}$ in the TIC1 model is attributed to the excitation from the $X$ state.
These results indicate that tunneling is the primary formation pathway, and excitation from the $X$ state is secondary.
On the other hand, the population at 0$^{\circ}$ in the TIC1 model is 43.3\% lower than that of the TIC1-D0 model.
This suggests that the population of the $B$ state is transferred to the $C$ state by their dipole coupling.

$C$ state --- The alignment dependence of the $C$ state population is similar to the $B$ state in both TIC1-D0 and TIC1 models.
The population at 0$^{\circ}$ in the TIC1 model is 51.6\% higher than that of the TIC1-D0 model.
These suggest that tunneling and excitation from the $B$ state are equally important to the $C$ state formation.
The enhancement at around 45$^{\circ}$ to 60$^{\circ}$ in the TIC1 model is attributed to the excitation from the $B$ state, which is formed by excitation from the $X$ state.

\begin{figure}
\begin{center}
\includegraphics[width=7.5cm]{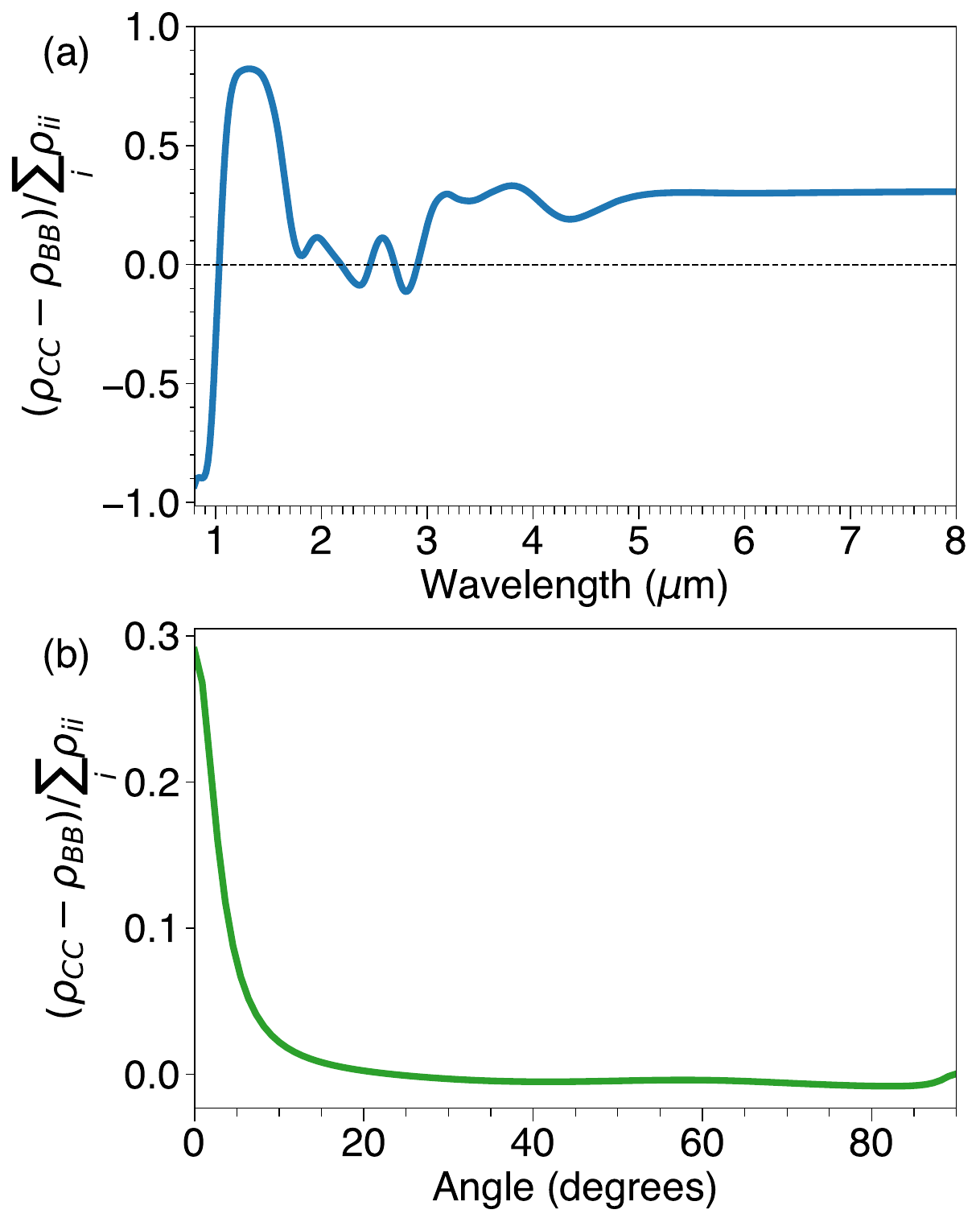}
\caption{(a) Wavelength dependence of the normalized population difference between the $C^{2}\Sigma_{g}^{+}$ and $B^{2}\Sigma_{u}^{+}$ states of CO$_{2}^{+}$ at 0$^{\circ}$ alignment with the laser polarization.
(b) Alignment dependence of the normalized population difference between the $C^{2}\Sigma_{g}^{+}$ and $B^{2}\Sigma_{u}^{+}$ at a wavelength of 5000 nm.
The results are calculated using the full DM-SFI theory with nonadiabatic rates (TIC1 model).
The laser pulses are 20 fs FWHM with a peak intensity of $1 \times 10^{14}$ W/cm$^{2}$.}
\label{fig:CO2-pop-inv}
\end{center}
\end{figure}

To further investigate the significant changes in the population of the $C$ state induced by the $CB$ dipole coupling, we perform wavelength scans ranging from 800 nm to 8000 nm, with a fixed pulse duration of 20 fs and a peak intensity of $1 \times 10^{14}$ W/cm$^{2}$.
Figure~\ref{fig:CO2-pop-inv}a displays the $CB$ population difference at 0$^{\circ}$ alignment, normalized by the total ionization yield.
We observe that the normalized $CB$ population difference varies rapidly from about $-0.9$ at 800 nm to 0.8 at 1300 nm.
Beyond 3200 nm, the normalized $CB$ population modulates weakly but remains inverted.
It becomes constant after 5000 nm.

\begin{figure*}
\begin{center}
\includegraphics[width=18cm]{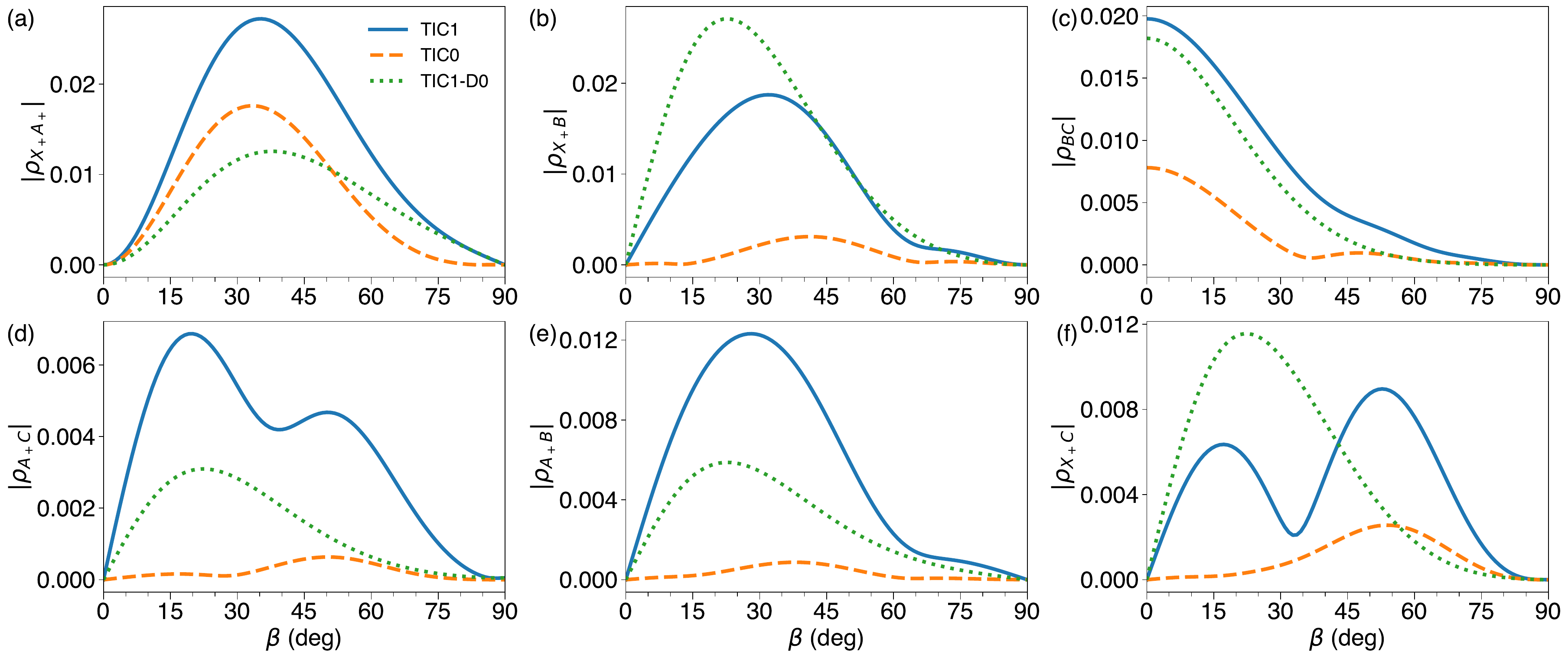}
\caption{Alignment dependence of the magnitude of ionic coherence between (a) $X^{2}\Pi_{g+}-A^{2}\Pi_{u+}$, (b) $X^{2}\Pi_{g+}-B^{2}\Sigma_{u}^{+}$, (c) $B^{2}\Sigma_{u}^{+}-C^{2}\Sigma_{g}^{+}$, (d) $A^{2}\Pi_{u+}-C^{2}\Sigma_{g}^{+}$, (e) $A^{2}\Pi_{u+}-B^{2}\Sigma_{u}^{+}$, and (f) $X^{2}\Pi_{g+}-C^{2}\Sigma_{g}^{+}$ states of CO$_{2}^{+}$ at the end of the pulse using the TIC1 (solid lines), TIC0 (dashed lines), and TIC1-D0 models (dotted lines). The laser parameters are identical to Fig.~\ref{fig:N2-subcycle}.}
\label{fig:CO2-ang-dist-coh}
\end{center}
\end{figure*}

Since populations of the other ionic states are zero at 0$^{\circ}$ alignment, the ion becomes a two-level system.
Consequently, the rapid changes at wavelengths shorter than 3200 nm can be explained by one- to multiphoton resonances resulting from ultrastrong couplings~\cite{Zhang2017, Chen2024, Li2025}.
At longer wavelengths, the population changes can be understood using adiabatic excitation theory~\cite{Yuen2025a}.
At 5000 nm, the second adiabaticity parameter is $\gamma_{e} = \omega/\Delta = 0.207$, where $\Delta = 1.2$ eV is the $CB$ energy separation, the excitation process becomes adiabatic~\cite{Yuen2025a}, and the normalized population difference becomes independent of wavelength.
Note that at higher peak intensities,  $\gamma_{e}$ must be even smaller to reach the adiabatic excitation regime so that the coupling of adiabatic states becomes negligible compared to the ionization rate.

While the analysis was done at 0$^{\circ}$ alignment, in Fig.~\ref{fig:CO2-pop-inv}b, the $CB$ population inversion persists until about 20$^{\circ}$ at 5000 nm.
Therefore, the $CB$ population inversion in the mid-infrared regime could be used for laser generation.
For example, CO$_2$ gas pumped by an intense mid-infrared laser pulse can serve as a gain medium for light with a frequency of approximately 1.2 eV.

\subsubsection{Ionic coherence~\label{sec6.2.2}}

\begin{table}
  \centering 
  \caption{Properties and the difference in the maximum magnitude of the ionic coherence relative to the TIC1 model. $P_{i} P_{j}$ is the product of the parity of the states. $\Delta E / \omega$ is the energy difference between the two states over the laser frequency (1.38 eV).}
  \begin{tabular}{ccccc}
\hline
\hline
State  & $P_{i} P_{j}$ & $\Delta E / \omega$  & TIC0 diff. & TIC1-D0 diff. \\
$X_{+}A_{+}$ & $-1$ & 2.82 & $-$35.4\% & $-$53.9\%  \\
$X_{+}B$ & $-1$ & 3.19 & $-$83.4\% & $+$44.7\%  \\
$BC$ & $-1$ & 0.870 & $-$60.5\% & $-$7.94\%  \\
$A_{+}C$ & $-1$ & 1.23 & $-$90.8\% & $-$55.0\%  \\
$A_{+}B$ & $+1$ & 0.362 & $-$92.9\% & $-$52.3\%  \\
$X_{+}C$ & $+1$ & 4.06 & $-$71.4\% & $+$28.9\%  \\
\hline
\end{tabular}
\label{tab:CO2-coh}
\end{table}

To understand the buildup of ionic coherence among different states of CO$_{2}^{+}$, in Fig.~\ref{fig:CO2-ang-dist-coh}, we compare the alignment dependence of the magnitude of the ionic coherence using the TIC1-D0, TIC0, and TIC1 models.
The coherence buildup mechanism is more complex than the population buildup because it involves two contributions: tunnel ionization coherences and dipole couplings.
The coherence in the TIC1-D0 and TIC0 models represents contributions from tunnel ionization coherences and dipole couplings, respectively.
In the following, we discuss each mechanism separately.

Tunnel ionization coherence becomes significant under specific conditions.
For states with opposite parity, $\Gamma_{ij}(t)$ in Eq.~\eqref{eq:coherence-rate} changes sign every half-cycle.
When the energy difference $E_{ij}$ is an odd multiple of the laser frequency $\omega$, then the coherence adds up constructively in each half-cycle; if $E_{ij}$ is an even multiple, the coherence adds out of phase.
The converse is true for states with the same parity.
Surprisingly, as shown in Table~\ref{tab:CO2-coh}, all coherences approximately satisfy the stated conditions for constructive interference.
The alignment dependence of the ionization rate also plays a role.
If the alignment distributions of the ionization rates do not significantly overlap, the tunnel ionization coherence will be diminished.
This effect can be seen in the results from the TIC1-D0 model in Fig.~\ref{fig:CO2-ang-dist-coh}: The alignment distributions resemble the product of the ionization rates in Fig.~\ref{fig:CO2-ang-dist}.

On the other hand, the coherence between states $i$ and $j$ builds from dipole couplings via the term 
\begin{align}
-i [\hat{H}(t), \hat{\rho}(t)]_{ij} = -i \sum_{k} [H_{ik} \rho_{kj} - \rho_{ik} H_{kj}]
\end{align}
in Eq.~\eqref{eq:EOM}.
The diagonal terms of the Hamiltonian give the quantum beat in the coherence.
If $H_{ij} \neq 0$, then $\rho_{ij}$ can build up from $-i H_{ij} (\rho_{jj} - \rho_{ii})$, so that it will scale with the population difference.
For example, $X_{+}A_{+}$, $X_{+}B$, and $BC$ coherences are predominantly driven by this term, as their alignment distribution in the TIC0 model in in Fig.~\ref{fig:CO2-ang-dist-coh} resembles the distribution of their population differences in Fig.~\ref{fig:CO2-ang-dist}.
In addition, $\rho_{ij}$ can build up from coherence between other states via $-i [H_{ik} \rho_{kj} - \rho_{ik} H_{kj}]$ for $k \neq i,j$.
For instance, in the TIC0 model, the $A_{+}C$ coherence can be seen as the average of $\rho_{X_{+}C}$ and $\rho_{A_{+}B}$ weighted by the dipole moment $\mathbf{d}_{A_{+}X_{+}} = 0.534 \, \mathbf{\hat{z}}$ and $\mathbf{d}_{BC} = 0.936 \, \mathbf{\hat{z}}$, respectively.
The $A_{+}B$ and $X_{+}C$ coherences in the TIC0 model can be analyzed in a similar manner.

Combining the buildup mechanisms from tunnel ionization coherences and dipole couplings significantly changes the peak value and the alignment distribution of the coherences.
From Table~\ref{tab:CO2-coh}, we observe that the peak coherences in the TIC0 model are 35.4 to 92.9\% smaller than those in the TIC1 model, while the peak coherences change from 7.94 to 55.0\% in the TIC1-D0 model relative to the TIC1 model.
The alignment distributions change notably for the $A_{+}C$ and $X_{+}C$ coherences.
In the TIC1 model, their distributions exhibit two maxima and a local minimum, with each peak corresponding to a peak in either the TIC0 or TIC1-D0 model.
We emphasize that the interplay between the two mechanisms is nonlinear, as tunnel ionization coherences can enhance dipole couplings. 

Ionic coherence is of significant importance for lasing and for driving electron motion in molecules.
As demonstrated in Gao \textit{et al.}~\cite{Gao2025co2}, the $XA$ ionic coherence leads to the strong emission of UV light.
Ionic coherences between other dipole-allowed pairs would also lead to the emission of light with a frequency around their energy separation.
For instance, the $XB$ and $BC$ coherences have comparable magnitudes and similar alignment dependence with the $XA$ coherence.
Therefore, the $XB$ and $BC$ coherences induce polarization in the gas medium parallel to the driving laser polarization, so lasing should also occur around 4.4 eV (282 nm) and 1.2 eV (1034 nm).
Given that the central wavelength of the considered driving laser is 900 nm, the $BC$ coherence may further amplify the driving laser.
On the other hand, the ionic coherence between dipole-allowed pairs could be probed using dissociative sequential double ionization spectroscopy~\cite{Yuen2024a, Weckwerth2025}.
The ionic coherence between dipole-forbidden pairs could be probed using attosecond transient absorption spectroscopy~\cite{Yuen2024c}.
Further experimental investigations could confirm the mechanisms of the coherence buildup in molecular ions created by strong laser fields.

\section{Conclusion~\label{sec7}}
To conclude, we developed a nonadiabatic theory of subcycle ionic dynamics in multielectron tunneling ionization.
The equivalence between the wave function and density matrix approaches for obtaining the reduced ionic density matrix is established.
We found that using the density matrix approach for subcycle ionic dynamics, which we call the DM-SFI theory, is advantageous because it avoids repeatedly solving the TDSE for the ion at different ionization times.
We also benchmarked the subcycle nonadiabatic ionization rate by comparing ionization yields calculated from the TDSE, confirming its quantitative accuracy.
Incorporating the nonadiabatic rates into the DM-SFI theory improves the quantitative accuracy of the reduced ionic density matrix compared to previous approaches~\cite{Yuen2023b, Yuen2024b}.

This work provides a theoretical foundation for density matrix approaches for strong field ionization and explains how population and coherence are built up in strong-field-generated molecular ions.
The validity of the DM-SFI theory also indicates that a recently proposed pump-control scheme for enhancing electronic coherence in an effective two-level molecular ion is promising~\cite{Yuen2025a}.
It should motivate future theoretical and experimental investigations into the control and observation of population and coherence in strong-field-generated molecular ions, thereby improving our understanding of multielectron tunneling and paving the way for its applications in lasing and chemical reaction control.
Additionally, the formal theory can be extended to sequential double ionization, which involves ionization from a superposition of ionic states.
The extended theory may explain the agreement between density matrix approaches~\cite{Yuen2022, Yuen2023, Jia2024, Jia2025} and experiments~\cite{Voss2004, Wu2010, Wu2010a}, and provide further insights into the mechanism of sequential double ionization.

\begin{acknowledgments}
This work was supported by the start-up fund provided by the College of Science and Mathematics at Kennesaw State University.
\end{acknowledgments}

\appendix
\section{Matching the subcycle ionization rate with cycle-averaged rate~\label{appendix1}}
To calculate the cycle-averaged ionization rate $w_{c}$, we start from Eq.~\eqref{eq:a_amp}.
Integrating over the momentum space, we have
\begin{align}
w_{c} &= \frac{\omega}{\pi} \int_{-\infty}^{\infty} \int_{-\infty}^{\infty} |a_{\mathbf{p}}(T/2)|^{2} dp_{\perp} dp_{z} \nonumber \\
 &\approx \frac{\omega}{\pi} \int_{-\infty}^{\infty} \int_{-\infty}^{\infty}  |C(t_{\mathbf{p}})|^{2} e^{2 \mathrm{Im}[S_{\mathbf{p}}(t_{\mathbf{p}})]} dp_{\perp} dp_{z}.
 \label{eq:T2yield}
\end{align}
The normalization factor $C$ can now be determined by matching the cycle-averaged Perelomov-Popov-Tenent'ev (PPT) rate~\cite{perelomov1966} with the Coulomb correction from Popruzhenko \textit{et al.}~\cite{Popruzhenko2008}.

The first step is to cast the exponent in Eq.~\eqref{eq:T2yield} as
\begin{align}
2 \mathrm{Im}[S_{\mathbf{p}}(t_{\mathbf{p}})] &= -\frac{2\kappa_{\perp}^{3}}{3F_{0}} g(p_{\perp}, t_{\mathbf{p}}),
\label{eq:recast}
\end{align}
with $\kappa_{\perp} = \sqrt{2E + p_{\perp}^{2}}$.
After some algebra, we find
\begin{align}
&g(p_{\perp}, t_{\mathbf{p}}) = \frac{3}{2\gamma_{K}} \left[ \left(1 + \sin^{2} \omega t_{\mathbf{p}} \frac{1 + \gamma_{\perp t_{\mathbf{p}}}^{2}}{\gamma^{2}_{K}} + \frac{1}{2\gamma_{K}^{2}}\right) \right.  \nonumber \\ 
&\times \operatorname{arcsinh} \gamma_{\perp t_{\mathbf{p}}} 
- \left. \frac{\gamma_{\perp t_{\mathbf{p}}} \sqrt{1 + \gamma_{\perp t_{\mathbf{p}}}^{2}}}{2 \gamma_{K}^{2}} (1 + 2\sin^{2}\omega t_{\mathbf{p}}) \right].
\label{eq:ggamma}
\end{align}
In particular,
\begin{align}
g(0, 0) = \frac{3}{2\gamma_{K}} \left[ \left(1 + \frac{1}{2\gamma_{K}^{2}}\right) \operatorname{arcsinh} \gamma_{K} - \frac{\sqrt{1 + \gamma^{2}_{K}}}{2 \gamma_{K}} \right]
\label{eq:g0}
\end{align}
coincides with the $g(\gamma)$ function in the PPT rate~\cite{perelomov1966}. For consistency with the previous theories, we define $g(\gamma_{K}) \equiv g(0,0)$.

Next, we apply the saddle point method to the integral of $p_{\perp}$. 
Since $g(p_{\perp}, t_{\mathbf{p}})$ changes much slower with respect to $p_{\perp}$ than $\kappa_{\perp}^{3}$ in Eq.~\eqref{eq:recast}, the exponent around $p_{\perp} =0$ can be well approximated as
\begin{align}
-\frac{2\kappa_{\perp}^{3}}{3F_{0}} g(p_{\perp}, t_{\mathbf{p}}) \approx -\frac{2\kappa^{3}}{3F_{0}} g(0, t_{p_{z}}) - \frac{\kappa}{F_{0}} g(0, t_{p_{z}}) p_{\perp}^{2},
\end{align}
with $\kappa = \sqrt{2E}$ and $t_{p_{z}} = t_{(0, p_z)}$.

Consequently, we have
\begin{align}
w_{c} &\approx \frac{\omega}{\pi} \int_{-\infty}^{\infty} \sqrt{\frac{\pi F_{0}}{\kappa g(0, t_{p_{z}})}} |C(t_{p_{z}})|^{2} \nonumber \\
& \times \exp{\left[-\frac{2\kappa^{3}}{3F_{0}} g(0, t_{p_{z}})\right]} dp_{z}.
\label{eq:p-perp-saddle}
\end{align}
To further abbreviate the notation, we denote $\gamma(0, t)$ in Eq.~\eqref{eq:gamma} as $\gamma_{t}$ and drop the $p_{\perp}$ dependence for $p_{z}$ in Eq.~\eqref{eq:pz} and for $g$ in Eq.~\eqref{eq:ggamma}.

We seek to apply the saddle point method again to the $p_{z}$ integral.
To do so, we first use
\begin{align}
\frac{dp_{z}}{dt} = |F(t)| \frac{1 + \gamma^{2}_{t}/\cos^{2} \omega t}{\sqrt{1 + \gamma^{2}_{t}}}
\label{eq:dpzdt}
\end{align}
to change the variable in the integral in Eq.~\eqref{eq:p-perp-saddle}.
The birth time $t_{p_{z}}$ is then replaced by $t$, and Eq.~\eqref{eq:p-perp-saddle} becomes
\begin{align}
w_{c}  &\approx \frac{1}{\pi} \int_{-\pi/2}^{\pi/2} \sqrt{\frac{\pi F_{0}}{\kappa g(t)}} |C(t)|^{2} \nonumber \\
& \exp{\left[-\frac{2\kappa^{3}}{3F_{0}} g(t)\right]} |F(t)| \frac{1 + \gamma^{2}_{t}/\cos^{2} \omega t}{\sqrt{1 + \gamma^{2}_{t}}} d \omega t,
\label{eq:b4-lastsaddle}
\end{align}
where we used $p_{z}(\pm \pi/2\omega) \to \pm \infty$.

The expansion of $g(t)$ around $t=0$ is
\begin{align}
g(t) \approx g(\gamma_{K}) + (\omega t)^{2} h(\gamma_{K}),
\end{align}
with
\begin{align}
h(\gamma_{K}) &= \frac{3}{4\gamma_{K}} \left[ 2 \left( 1+ \frac{1}{\gamma_{K}^2} \right) \operatorname{arcsinh} \gamma_{K} \right. \nonumber \\
& \left. -\frac{3\sqrt{1 + \gamma^{2}_K}}{\gamma_{K}} +  \left( 1+ \frac{1}{\gamma_{K}^2} \right) \frac{\gamma_{K}}{\sqrt{1 + \gamma^{2}_K}} \right].
\label{eq:hfunc}
\end{align}

Applying the saddle point method, Eq.~\eqref{eq:b4-lastsaddle} becomes
\begin{align}
w_{c}  &\approx |C(t=0)|^{2} \frac{F^{2}_{0}}{\kappa^{2}} \sqrt{\frac{3 (1 + \gamma^{2}_{K})}{ 2 g(\gamma_{K}) h(\gamma_{K})}}  \exp{\left[-\frac{2\kappa^{3}}{3F_{0}} g(\gamma_{K}) \right]}
\label{eq:lastsaddle}
\end{align}

Matching the improved PPT rate~\cite{perelomov1966, Popruzhenko2008} for an orbital with a magnetic quantum number $m$, we find
\begin{align}
|C(t=0)|^{2} &= \frac{|B_{m}|^{2}}{2^{|m|} |m|!} \frac{1}{F_{0} \kappa^{2Z/\kappa}} 
\sqrt{\frac{g(\gamma_{K}) h(\gamma_{K})}{\pi}} \nonumber \\
& \times A_{m}(\omega, \gamma_{K}) 
\left(\frac{F_{0} \sqrt{1+\gamma_{K}^{2}}}{2 \kappa^{3}}\right)^{|m| + \frac{1}{2}}  \nonumber \\
& \times \left( \frac{2\kappa^{3}}{F_{0}} \right)^{2Z/\kappa}
\left( 1 + 2 e^{-1} \gamma_{K} \right)^{-2Z/\kappa}.
\label{eq:normalization-full}
\end{align}

In the above, we have~\cite{Yuen2024b}
\begin{align}
B_{m} &= \sum_{lm'} \mathrm{sgn}\left[-F_{0}\right]^{l-m} C_{lm'}  D^l_{m m'}(\hat{R}) Q(l,m), \label{eq:Bm} \\
Q(l,m) &= (-1)^{(m + |m|)/2} \sqrt{\frac{2l+1}{2} \frac{(l+|m|)!}{(l-|m|)!}},
\end{align}
where $\mathrm{sgn}\left[-F_{0}\right]= \pm 1$ for $F_{0}<0$ or $F_{0}>0$, $C_{lm}$ is the structure parameter of the orbital~\cite{Tong2002, Zhao2010, Zhao2011}, and $D^l_{m m'}$ is the Wigner $D$ matrix, where the $z-y-z$ extrinsic rotation convention~\cite{Morrison1987} is used for the Euler angles $\hat{R}$.  

The $A_{m}(\omega, \gamma_{K})$ function is given by~\cite{perelomov1966}
\begin{align}
A_{m}(\omega, \gamma_{K}) &= \frac{4}{3\pi} \frac{1}{|m|!} \frac{\gamma_{K}^{2}}{1 + \gamma_{K}^{2}} \nonumber \\
& \times \sum_{n \geq \nu} e^{-\alpha (n- \nu)} w_{m}(\sqrt{\beta(n - \nu)}), \label{eq:Am} \\
w_{m}(x) &= \frac{x^{2|m|+1}}{2} \int_{0}^{1} \frac{e^{-x^{2}t}t^{|m|}}{\sqrt{1-t}} dt, \\
\alpha &= 2 \left[ \operatorname{arcsinh} \gamma_{K} - \frac{\gamma_{K}}{\sqrt{1 + \gamma_{K}^{2}}}\right], \\
\beta &= 2\gamma_{K} / \sqrt{1 + \gamma_{K}^{2}}, \\
\nu &= \frac{E}{\omega} \left( 1 + \frac{1}{2\gamma_{K}^{2}}\right).
\end{align}
Note that when $m=0$, $w_{0}(x)$ reduces to the Dawson integral with an analytical expression,
\begin{align}
w_{0}(x) = \frac{\sqrt{\pi}}{2} e^{-x^{2}} \mathrm{erfi}(x),
\end{align}
where $\mathrm{erfi}$ is the imaginary error function.

Finally, to obtain the subcycle ionization rate, we first calculate the instantaneous ionization yield at time $t$.
Similar to Eq.~\eqref{eq:b4-lastsaddle}, it is
\begin{align}
\rho(t) &=  \int_{-\frac{\pi}{2\omega}}^{t} \sqrt{\frac{\pi F_{0}}{\kappa g(t_{b})}} |C(t_{b})|^{2} 
 \exp{\left[-\frac{2\kappa^{3}}{3F_{0}} g(t_{b})\right]} \nonumber \\
& \times |F(t_{b})| \frac{1 + \gamma^{2}_{t_{b}}/\cos^{2} \omega t_{b}}{\sqrt{1 + \gamma^{2}_{t_{b}}}} d t_{b}.
\end{align}

Then, the subcycle ionization rate $w(t) = d \rho/dt$ can be identified as
\begin{align}
w(t) &= |C(t)|^{2} \sqrt{\frac{\pi F_{0}}{\kappa g(t)}} |F(t)| \frac{1 + \gamma^{2}_{t}/\cos^{2} \omega t}{\sqrt{1 + \gamma^{2}_{t}}} \nonumber \\
&\exp{\left[-\frac{2\kappa^{3}}{3F_{0}} g(t)\right]}.
\label{eq:rate2compare}
\end{align}

Approximating all the prefactors of the exponential function at $t=0$,  the subcycle for an orbital with a magnetic quantum number $m$ is then
\begin{align}
w_{m}(t) &= \frac{|B_{m}|^{2}}{2^{|m|} |m|!} 
\frac{1}{\kappa^{2Z/\kappa-1}} 
\left( 2h(\gamma_{K})\sqrt{1 + \gamma_{K}^{2}} \right)^{1/2} \nonumber \\
& \times  A_{m}(\omega, \gamma_{K}) 
\left(\frac{F_{0} \sqrt{1+\gamma_{K}^{2}}}{2 \kappa^{3}}\right)^{|m| + 1}  \nonumber \\
& \times  \left( \frac{2\kappa^{3}}{F_{0}} \right)^{2Z/\kappa}
\left( 1 + 2 e^{-1} \gamma_{K} \right)^{-2Z/\kappa}
\exp{\left[-\frac{2\kappa^{3}}{3F_{0}} g(t)\right]}.
\end{align}

\section{Derivation of the equation of motion for the reduced ionic density matrix~\label{appendix2}}
To derive Eq.~\eqref{eq:coherence-rate} in the main text, we start by writing
\begin{align}
\Gamma_{ij}(t) &=  \int d\mathbf{p} \, \rho_0 (t_{\mathbf{p}}) \sum_{i,j} 2 C_{i}(t_{\mathbf{p}}) C^{\ast}_{j}(t_{\mathbf{p}}) \nonumber \\
&\times  \delta(t - t_{\mathbf{p}}) \Theta(t - t_{\mathbf{p}}) e^{\mathrm{Im} \left[S_{i \mathbf{p}}(t_{\mathbf{p}}) + S_{j \mathbf{p}}(t_{\mathbf{p}})\right]}.
\end{align}

Following Eqs.~\eqref{eq:recast} and \eqref{eq:ggamma}, we cast the exponents as
\begin{align}
\mathrm{Im} \left[S_{i \mathbf{p}}(t_{\mathbf{p}}) \right] = \frac{\kappa_{i \perp}^{3}}{3F_{0}} g_{i}(p_{\perp}, t),
\end{align}
where the subscript $i$ means the ionization potential $E$ is replaced by $E_{i}$.

Next, we apply the saddle point method for the $p_{\perp}$ integral similar to Eq.~\eqref{eq:p-perp-saddle}, and obtain
\begin{align}
\Gamma_{ij}(t) &= \int_{-\infty}^{\infty} dp_{z} \sqrt{\frac{2\pi F_{0}}{\kappa_{i} g_{i}(t_{p_{z}}) + \kappa_{j} g_{j}(t_{p_{z}})}} C_{i}(t_{p_{z}}) C^{\ast}_{j}(t_{p_{z}}) \nonumber \\
&\times 2 \rho_0 (t_{p_{z}}) \delta(t - t_{p_{z}}) \Theta(t - t_{p_{z}}) \nonumber \\
& \times \exp{\left[ - \frac{\kappa_{i}^{3} g_{i}(t_{p_{z}}) + \kappa_{j}^{3} g_{j}(t_{p_{z}})}{3F_{0}}\right]},
\end{align}
where $\mathbf{p} = p_{z}$ as $p_{\perp} = 0$. 

Next, using Eq.~\eqref{eq:dpzdt}, we transform the $p_{z}$ integral to 
\begin{align}
\Gamma_{ij}(t) &= \int_{-\pi/2\omega}^{\pi/2\omega} dt_{p_{z}} \, 
|F(t_{p_{z}})| \frac{1 + \gamma^{2}_{t_{p_{z}}}/\cos^{2} \omega t_{p_{z}}}{\sqrt{1 + \gamma^{2}_{t_{p_{z}}}}}  \nonumber \\ 
&\times \sqrt{\frac{2\pi F_{0}}{\kappa_{i} g_{i}(t_{p_{z}}) + \kappa_{j} g_{j}(t_{p_{z}})}} 
\nonumber \\ 
&\times 2\rho_0 (t_{p_{z}}) C_{i}(t_{p_{z}}) C^{\ast}_{j}(t_{p_{z}})  \delta(t - t_{p_{z}}) \Theta(t - t_{p_{z}}) \nonumber \\
& \times \exp{\left[ - \frac{\kappa_{i}^{3} g_{i}(t_{p_{z}}) + \kappa_{j}^{3} g_{j}(t_{p_{z}})}{3F_{0}}\right]}.
\end{align}
We contract the time integral and have
\begin{align}
\Gamma_{ij}(t) &= \rho_0 (t) C_{i}(t) C^{\ast}_{j}(t) \sqrt{\frac{2\pi F_{0}}{\kappa_{i} g_{i}(t) + \kappa_{j} g_{j}(t)}}  \nonumber \\ 
&\times |F(t)| \frac{1 + \gamma^{2}_{t}/\cos^{2} \omega t}{\sqrt{1 + \gamma^{2}_{t}}} \exp{\left[ - \frac{\kappa_{i}^{3} g_{i}(t) + \kappa_{j}^{3} g_{j}(t)}{3F_{0}}\right]},
\end{align}
 where we used $\Theta(0) = 1/2$.
 To cast $\Gamma_{ij}$ in terms of ionization rates, we first approximate 
\begin{align}
\sqrt{\frac{\kappa_{i} g_{i}(t) + \kappa_{j} g_{j}(t)}{2}} \approx
\left[\kappa_{i} g_{i}(t) \kappa_{j} g_{j}(t) \right]^{1/4},
\end{align}
which says that the arithmetic mean of $\kappa_{i} g_{i}(t)$ and $\kappa_{j} g_{j}(t)$ is nearly equal to its geometric mean.
For orbitals 1 and 2 in Sec.~\ref {sec4}, these two means differ by less than 0.1\%. 
Then, we write
\begin{align}
\Gamma_{ij}(t) &= \rho_0 (t) \frac{C_{i}(t) C^{\ast}_{j}(t)}{|C_{i}(t)| |C_{j}(t)|}  |F(t)| \frac{1 + \gamma^{2}_{t}/\cos^{2} \omega t}{\sqrt{1 + \gamma^{2}_{t}}} \nonumber \\ 
&\times |C_{i}(t)| \left[ \frac{\pi F_{0}}{\kappa_{i} g_{i}(t)} \right]^{1/4}  \exp{\left[ - \frac{\kappa_{i}^{3} g_{i}(t)}{3F_{0}}\right]} \nonumber \\ 
&\times |C_{j}(t)| \left[ \frac{\pi F_{0}}{\kappa_{j} g_{j}(t)} \right]^{1/4} \exp{\left[ - \frac{\kappa_{j}^{3} g_{j}(t)}{3F_{0}}\right]}.
\label{eq:gammaij-last}
\end{align}
In the normalization factor $C_{i}$~\eqref{eq:normalization-full}, only the structure term $B_{m}$~\eqref{eq:Bm} is complex since the Coulomb phase cancelled out.
Therefore, comparing Eq.~\eqref{eq:gammaij-last} with Eq.~\eqref{eq:rate2compare}, we have Eq.~\eqref{eq:coherence-rate}.

\section{Computational details~\label{appendix3}}

To solve the reduced ionic density matrix $\rho_{N}(t)$ for a Gaussian pulse with a FWHM $\tau$, the time grid was set to $-\tau$ to $2 \tau$ with a step of 10 as.
We first interpolated the population of the neutral state, Eq.~\eqref{eq:pop0}.
Then, setting $\rho_{N}(t_{0}) = 0$, we solve Eq.~\eqref{eq:EOM} with Eq.~\eqref{eq:coherence-rate} using the explicit Runge-Kutta method of order 5(4).
To control solution accuracy, the relative and absolute tolerances were set to $10^{-8}$ and $10^{-10}$, respectively, and the maximum time step was set to 0.1 fs.

To determine the structure parameters of N$_{2}$ and  CO$_{2}$, we employed the real-space density functional theory software \texttt{Octopus}~\cite{TancogneDejean2020}.
The asymptotic behavior for the Coulomb potential is properly imposed for Kohn-Sham orbitals using the corrected exchange density with the local density approximation (LDA)~\cite{Andrade2011}.
For all molecules, the molecular axis was taken to be the $z$-axis, with the origin at the center of mass.
Since \texttt{Octopus} does not impose symmetry, a symmetrization procedure was used to transform the calculated $\pi$ orbitals into $\pi_{\pm}$.
The structure parameters were extracted at the radius where the magnitude of the radial wave function is 10$^{-7}$.
The structure parameter with the largest magnitude is set to be real and positive.
Table~\ref{Tab:Clm} presents the structure parameters of the orbitals of N$_{2}$ and CO$_{2}$ used in this work.
Below are the details involved for each molecules.

\begin{table}[t]
 \caption{Structure parameters $C_{lm}$ and ionization potentials of the orbitals of N$_{2}$ and CO$_{2}$ used in this work.}
\centering
\begin{tabular}{p{2.0cm}p{2.0cm}p{4cm}}
\hline
\hline
Orbital & $I_p$ (eV) & $C_{lm}$\\
N$_{2}$, $3\sigma_{g}$ & 15.6 &  
\begin{tabular}{ccc}
$C_{00}$ & $C_{20}$ & $C_{30}$\\
$3.16$ & $1.15$ & $0.06$ \\
\end{tabular}\\
N$_{2}$, $1\pi_{u\pm}$ & 16.9 &  
\begin{tabular}{cc}
$C_{1\pm1}$ &   $C_{3\pm1}$ \\
$2.25$ & $0.24$ \\
\end{tabular}\\
N$_{2}$, $2\sigma_{u}$ & 18.8 & \begin{tabular}{cc}
$C_{10}$ &  $C_{30}$  \\
$4.16$ & $0.35$ \\
\end{tabular}\\
\\
CO$_{2}$, $1\pi_{g\pm}$ & 13.8 &  
\begin{tabular}{cc}
$C_{2\pm 1}$ & $C_{4 \pm 1}$ \\
$1.62$ & $0.29$  \\
\end{tabular}\\
CO$_{2}$, $1\pi_{u\pm}$ & 17.7 &  
\begin{tabular}{ccc}
$C_{1\pm1}$ &   $C_{3\pm1}$ &   $C_{5\pm1}$\\
$3.17$ & $1.06$ & $0.15$ \\
\end{tabular}\\
CO$_{2}$, $3\sigma_{u}$ & 18.2 & 
\begin{tabular}{ccc}
$C_{10}$ &   $C_{30}$ &   $C_{50}$\\
$6.81$ & $2.38$ & $0.3$ \\
\end{tabular}\\
CO$_{2}$, $4\sigma_{g}$ & 19.4 & 
\begin{tabular}{cccc}
$C_{00}$ &  $C_{20}$ &  $C_{40}$ &  $C_{60}$\\
$5.46$ & $5.61$ & $1.07$ & $0.1$\\
\end{tabular}\\
\hline
\hline
\end{tabular}
\label{Tab:Clm}
\end{table}

\subsection{N$_{2}$~\label{appendix3.1}}
The bond length of N$_{2}$ is set to 1.10 \AA.
The simulation box is a sphere with radius 30 $a_{0}$ and a spacing of 0.3 $a_{0}$.
The LDA norm-conserving pseudo-potential was used.

\subsection{CO$_{2}$~\label{appendix3.2}}
The CO bond length is set to 1.16 \AA.
The simulation box is a sphere with radius 20 $a_{0}$ and a spacing of 0.3 $a_{0}$.
The LDA norm-conserving pseudo-potential was used.

The transition dipole moments between different CO$_{2}^{+}$ states were calculated using the software \texttt{OpenMolcas}~\cite{Fdez.Galvan2019} using the state-averaged complete active space self-consistent field method and the ANO-L basis.
The $C_{1}$ symmetry group was used so that the electronic wave functions with different symmetries
are optimized on an equal footing.
The lowest five orbitals were frozen, and 11 active electrons were put into 10 active orbitals.
Six electronic states were included in the state-averaged calculations. 
The second-order perturbation (CASPT2) calculation was performed to improve further the accuracy of the
electronic energies for state identification.

\end{document}